\documentclass[superscriptaddress,groupedaddress,nofootinbib,preprintnumbers,12pt]{revtex4-1} 
\usepackage{amsmath}

\usepackage{graphicx}        
\usepackage[bottom]{footmisc}

\usepackage{newtxtext}       %
\usepackage[varvw]{newtxmath}       

\makeindex             

\RequirePackage[normalem]{ulem} 
\RequirePackage{color}\definecolor{RED}{rgb}{1,0,0}\definecolor{BLUE}{rgb}{0,0,1} 
\providecommand{\DIFaddbegin}{} 
\providecommand{\DIFaddend}{} 
\providecommand{\DIFdelbegin}{} 
\providecommand{\DIFdelend}{} 
\providecommand{\DIFaddbeginFL}{} 
\providecommand{\DIFaddendFL}{} 
\providecommand{\DIFdelbeginFL}{} 
\providecommand{\DIFdelendFL}{} 
\newcommand{\DIFscaledelfig}{0.5}
\RequirePackage{settobox} 
\RequirePackage{letltxmacro} 
\newsavebox{\DIFdelgraphicsbox} 
\newlength{\DIFdelgraphicswidth} 
\newlength{\DIFdelgraphicsheight} 
\LetLtxMacro{\DIFOincludegraphics}{\includegraphics} 
\newcommand{\DIFaddincludegraphics}[2][]{{\color{blue}\fbox{\DIFOincludegraphics[#1]{#2}}}} 
\newcommand{\DIFdelincludegraphics}[2][]{
\sbox{\DIFdelgraphicsbox}{\DIFOincludegraphics[#1]{#2}}
\settoboxwidth{\DIFdelgraphicswidth}{\DIFdelgraphicsbox} 
\settoboxtotalheight{\DIFdelgraphicsheight}{\DIFdelgraphicsbox} 
\scalebox{\DIFscaledelfig}{
\parbox[b]{\DIFdelgraphicswidth}{\usebox{\DIFdelgraphicsbox}\\[-\baselineskip] \rule{\DIFdelgraphicswidth}{0em}}\llap{\resizebox{\DIFdelgraphicswidth}{\DIFdelgraphicsheight}{
\setlength{\unitlength}{\DIFdelgraphicswidth}
\begin{picture}(1,1)
\thicklines\linethickness{2pt} 
{\color[rgb]{1,0,0}\put(0,0){\framebox(1,1){}}}
{\color[rgb]{1,0,0}\put(0,0){\line( 1,1){1}}}
{\color[rgb]{1,0,0}\put(0,1){\line(1,-1){1}}}
\end{picture}
}\hspace*{3pt}}} 
} 
\LetLtxMacro{\DIFOaddbegin}{\DIFaddbegin} 
\LetLtxMacro{\DIFOaddend}{\DIFaddend} 
\LetLtxMacro{\DIFOdelbegin}{\DIFdelbegin} 
\LetLtxMacro{\DIFOdelend}{\DIFdelend} 
\DeclareRobustCommand{\DIFaddbegin}{\DIFOaddbegin \let\includegraphics\DIFaddincludegraphics} 
\DeclareRobustCommand{\DIFaddend}{\DIFOaddend \let\includegraphics\DIFOincludegraphics} 
\DeclareRobustCommand{\DIFdelbegin}{\DIFOdelbegin \let\includegraphics\DIFdelincludegraphics} 
\DeclareRobustCommand{\DIFdelend}{\DIFOaddend \let\includegraphics\DIFOincludegraphics} 
\LetLtxMacro{\DIFOaddbeginFL}{\DIFaddbeginFL} 
\LetLtxMacro{\DIFOaddendFL}{\DIFaddendFL} 
\LetLtxMacro{\DIFOdelbeginFL}{\DIFdelbeginFL} 
\LetLtxMacro{\DIFOdelendFL}{\DIFdelendFL} 
\DeclareRobustCommand{\DIFaddbeginFL}{\DIFOaddbeginFL \let\includegraphics\DIFaddincludegraphics} 
\DeclareRobustCommand{\DIFaddendFL}{\DIFOaddendFL \let\includegraphics\DIFOincludegraphics} 
\DeclareRobustCommand{\DIFdelbeginFL}{\DIFOdelbeginFL \let\includegraphics\DIFdelincludegraphics} 
\DeclareRobustCommand{\DIFdelendFL}{\DIFOaddendFL \let\includegraphics\DIFOincludegraphics} 

\begin{document}

\title{Non-Gaussianities in primordial black hole formation and induced gravitational waves}
\author{Shi Pi${}^{a,b,c}$} \email{shi.pi@itp.ac.cn}		
	\affiliation{
	$^{a}$ Institute of Theoretical Physics, Chinese Academy of Sciences, Beijing 100190, China\\
		$^{b}$ Center for High Energy Physics, Peking University,  Beijing 100871, China\\
		$^{c}$ Kavli Institute for the Physics and Mathematics of the Universe (WPI), The University of Tokyo, Kashiwa, Chiba 277-8583, Japan}
	\date{\today}

\begin{abstract}
The most promising mechanism of generating PBHs is by the enhancement of power spectrum of the primordial curvature perturbation, which is usually accompanied by the the enhancement of non-Gaussianity that crucially changes the abundance of PBHs. In this review I will discuss how non-Gaussianity is generated in single field inflation as well as in the curvaton scenario, and then discuss how to calculate PBH mass function and induced gravitational waves (GWs) with such non-Gaussianities. 
When the PBH abundance is fixed, non-Gaussianity only has mild impact on the spectral shape of the induced GWs, which gives relatively robust predictions in the mHz and nHz GW experiments.
\end{abstract}
\maketitle

\section{Overview}

PBHs are formed when the density contrast $\delta\rho/\rho$ exceeds some critical value $\delta_\mathrm{th}\sim0.4$ when the perturbation reenter the Hubble horizon \cite{Zeldovich:1967lct,Hawking:1971ei,Carr:1974nx,Meszaros:1974tb,Carr:1975qj,Khlopov:1985jw}. The peak mass of PBHs depends on the horizon mass at the reentry moment of the over-densed region, while the abundance of the PBHs depends on the number of such Hubble patches which can satisfy the condition of PBH formation (like $\delta\rho/\rho>\delta_\mathrm{th}$). Motivated by the Press-Schechter formalism in calculating the halo mass function in galaxy formation \cite{Press:1973iz}, the PBH abundance was calculated by integrating the probability density function (PDF) of the density contrast $\mathbb{P}(\delta\rho/\rho)$ from the threshold. Later, it was realized that the criterion for PBH formation should be set on the compaction function $\mathscr{C}(t,r)= 2G\delta M(t,r)/r$ \cite{Shibata:1999zs}, which is a direct application of the hoop conjecture \cite{Misner:1973prb}. 
The calculation based on the PDF of compaction function $\mathbb{\mathscr{C}}$ is called the Press-Schechter-type method, which is reviewed in Chapter 7 \cite{Young:2024jsu} under the assumption that the curvature perturbation is Gaussian. 
By the observation of the cosmic microwave background (CMB) anisotropies and the large scale structure, we already confirm the primordial curvature perturbation on comoving slices, $\mathcal{R}$, is nearly scale invariant and Gaussian, with a power spectrum $\mathcal{P_R}\sim10^{-9}$ on scales larger than about $1~\text{Mpc}$~\cite{Akrami:2018odb}. However, this is not enough to generate detectable amount of PBHs (say, as microlensing events, black hole merger events, with detectable stochastic GW from the mergers, scalar induced GW, \textit{etc}.), and a rough estimate show that the power spectrum of the curvature perturbation should be enhanced to $\mathcal{P_R}\sim10^{-2}$. If it is still Gaussian, the number of Hubble patches of PBH formation is solely determined by the power spectrum and the threshold of the compaction function \footnote{Besides the variance and the threshold, the PBH abundance also depends on the choice of window functions. This is especially important when the power spectrum is broad. We ignore this dependence for simplicity.}. 

A related phenomenon is the stochastic gravitational wave background induced by the enhanced curvature perturbation. In General Relativity, the curvature perturbation $\mathcal{R}$ and tensor perturbation $h_{ij}$ are only coupled at nonlinear order, and the tensor-scalar-scalar type coupling $\sim h_{ij}\partial_i\mathcal{R}\partial_j\mathcal{R}$ can induce second-order GWs \cite{Matarrese:1992rp,Matarrese:1993zf,Matarrese:1997ay,Noh:2004bc,Carbone:2004iv,Nakamura:2004rm,Ananda:2006af,Osano:2006ew,Baumann:2007zm} \footnote{For GWs induced by other couplings like scalar-tensor-tensor or tensor-tensor-tensor, see for instance Refs.~\cite{Chang:2022vlv,Yu:2023lmo,Bari:2023rcw,Picard:2023sbz}.}. Therefore, the curvature perturbation can source the GWs after it reenters the Hubble horizon, which is called scalar induced secondary GWs. The peak frequency, as well as the PBH mass generated by such a perturbation, are determined by the horizon scale at the reentry moment, which are connected by \cite{Saito:2008jc}
\begin{equation}\label{eqn8:fgw-pbhmass}
f_\mathrm{IGW}\approx3\mathrm{Hz}\left(\frac{M_\mathrm{PBH}}{10^{16}\mathrm{g}}\right)^{-\frac{1}{2}}.
\end{equation}
At the peak scale, the spectral energy density of the induced GWs we observe today (normalized by the critical density), is roughly $\Omega_\text{GW,0}\equiv\rho_\mathrm{cr}^{-1}\left(\mathrm{d}\rho_\mathrm{GW}/\mathrm{d}\ln k\right)_{t_0}\sim10^{-6}\mathcal{P}_\mathcal{R}^2$, where $\mathcal{P}_\mathcal{R}$ is the primordial power spectrum of $\mathcal{R}$. The nearly scale-invariant power spectrum extrapolated from CMB scales predicts a scale-invariant induced GW with an amplitude of $\Omega_\text{GW,0}\sim10^{-24}$, far less than observational bound \cite{Baumann:2007zm}. However, for the enhanced curvature perturbation with abundant PBH formation ($\mathcal{P_R}\sim\mathcal{O}(10^{-2})$), the induced GW is also enhanced to $\Omega_\text{GW,0}\sim10^{-10}$, which reaches the detectable range of many experiments like the next-generation ground- and space-borne interferometers and pulsar timing arrays. As both $\Omega_\text{GW,0}$ and PBH abundance $f_\mathrm{PBH}$ depend on $\mathcal{P_R}$, they can be cross-checked to probe the primordial power spectrum on small scales~\cite{Saito:2008jc,Saito:2009jt,Bugaev:2009zh,Assadullahi:2009jc,Bugaev:2010bb}. 
For more discussion on the induced GWs, see Chapter 18 \cite{Domenech:2024kmh}.

To enhance the power spectrum of the curvature perturbation, usually a deviation from the slow-roll attractor is required. Depending on the model, this not only changes the shape and amplitude of the enhanced power spectrum, but also the statistics of the curvature perturbation.
Even a small non-Gaussianity can change the relation between the PBH abundance and induced GW spectrum significantly, as the induced GW depends mainly on the variance of the PDF of the curvature perturbation, while the PBH abundance depends on the high-$\sigma$ tail of the PDF. A small change in the PDF, \textit{i.e.} deviation from the Gaussianity, can give rise to a huge enhancement/suppression in the PBH abundance \cite{Bullock:1996at,Ivanov:1997ia,Yokoyama:1998xd,PinaAvelino:2005rm,Seery:2006wk,Hidalgo:2007vk,Byrnes:2012yx,Young:2013oia,Young:2015cyn,Atal:2018neu,Yoo:2019pma,Kehagias:2019eil,Mahbub:2020row,Riccardi:2021rlf,Taoso:2021uvl,Biagetti:2021eep,Kitajima:2021fpq,Young:2022phe,Escriva:2022pnz,Matsubara:2022nbr,Gow:2022jfb}, but only changes the amplitude of the induced GW spectrum mildly \cite{Nakama:2016gzw,Garcia-Bellido:2017aan,Cai:2018dig,Unal:2018yaa,Unal:2020mts,Adshead:2021hnm,Garcia-Saenz:2022tzu,Abe:2022xur}. 

On large scales, non-Gaussianity is often described by the perturbative series \cite{Komatsu:2001rj}
\begin{equation}\label{eqn8:Rseries}
\mathcal{R}=\mathcal{R}_g+\frac{3}{5}f_\mathrm{NL}(\mathcal{R}_g^2-\langle\mathcal{R}_g\rangle^2)+\cdots,
\end{equation}
which is also used to describe the curvature perturbation on small scales for PBH formation \cite{Byrnes:2012yx,Young:2013oia,Tada:2015noa,Young:2015kda,Young:2015cyn,Franciolini:2018vbk,Ando:2018nge,Atal:2018neu,Passaglia:2018ixg,Ozsoy:2023ryl} as well as the induced GWs \cite{Nakama:2016gzw,Garcia-Bellido:2017aan,Cai:2018dig,Unal:2018yaa,Unal:2020mts}. However, in many models, the non-linear parameter $f_\mathrm{NL}$ reaches $\mathcal{O}(1)$ \cite{Namjoo:2012aa,Martin:2012pe,Chen:2013aj,Motohashi:2014ppa,Davies:2021loj,Namjoo:2023rhq,Namjoo:2024ufv}, so the higher orders in \eqref{eqn8:Rseries} might be important, which are highly model-dependent. These higher order terms must be taken into account in the PBH formation. Fortunately, by $\delta N$ formalism, a fully nonlinear curvature perturbation can be derived for single field inflation and for the curvaton scenario, of which the PDF of $\mathcal{R}$ can be calculated analytically. Various methods are developed to calculate the PBH abundance for such a PDF, including the Press-Schechter-type formalism \cite{Biagetti:2021eep,Gow:2022jfb,Ferrante:2022mui} and the theory of peaks \cite{Green:2004wb,Yoo:2018kvb,Germani:2018jgr,Atal:2019cdz,Atal:2019erb,Young:2020xmk,Yoo:2020dkz,Taoso:2021uvl,Riccardi:2021rlf,Kitajima:2021fpq,Young:2022phe}. The Press-Schechter-type formalism for Gaussian curvature perturbation is thoroughly reviewed in Chapter 6 \cite{Young:2024jsu}, which is 
straightforward to be further extended to non-Gaussian case. 

Considering non-Gaussianity in the PBH formation and induced GW generation is important in many theoretical and observational issues in cosmology. 
\begin{itemize}
\item
PBHs might be the supermassive or stupendously large BHs which seed the galaxy or structure formation~\cite{Bean:2002kx,Kawasaki:2012kn,Carr:2018rid,Carr:2020erq}. To satisfy the stringent constraints from CMB $\mu$- and $y$-distortion \cite{Kohri:2014lza}, large non-Gaussianities must be included \cite{Nakama:2016kfq,Nakama:2017xvq,Nakama:2019htb,Atal:2020yic,Carr:2018rid,Liu:2022bvr,Biagetti:2022ode,Gouttenoire:2023nzr,Hooper:2023nnl}. 
\item
According to the current observational constraints, asteroid-mass PBH can be all the dark matter, of which the accompanying induced GWs are detectable by space-borne interferometers LISA \cite{Barausse:2020rsu,LISA:2022kgy,LISACosmologyWorkingGroup:2022jok,LISACosmologyWorkingGroup:2023njw}, Taiji \cite{Ren:2023yec}, and TianQin \cite{Liang:2021bde}. As positive non-Gaussianity can enhance the PBH abundance greatly, the required power spectrum of the curvature perturbation for $\Omega_\mathrm{PBH}=\Omega_\mathrm{DM}$ is reduced with positive non-Gaussianity, and the induced GW spectrum is also reduced, yet it is still above the LISA/Taiji/TianQin sensitivity curve \cite{Cai:2018dig,Bartolo:2018evs,Unal:2018yaa}. The large non-Gaussian limit with $f_\mathrm{NL}\gg1$ (\textit{i.e.} the $\chi^2$-distribution) can be realized in the curvaton model when curvaton is negligible at its decay \cite{Bartolo:2003jx,Enqvist:2005pg,Sasaki:2006kq,Pi:2021dft,Ferrante:2022mui}. 
\item
In 2023, the Pulsar Timing Array collaborations NANOGrav~\cite{NANOGrav:2023gor,NANOGrav:2023hde}, EPTA combined with InPTA~\cite{EPTA:2023fyk,EPTA:2023sfo,EPTA:2023xxk}, PPTA\,\cite{Zic:2023gta,Reardon:2023gzh,Reardon:2023zen}, CPTA\,\cite{Xu:2023wog}, as well as IPTA \cite{InternationalPulsarTimingArray:2023mzf}, announced the strong evidence of nHz stochastic GW background, while the induced GW interpretation is favored \cite{NANOGrav:2023hvm}. However, as the amplitude of the nHz GW spectrum is relatively high ($\sim10^{-9}$), sub-solar-mass PBHs might be overproduced, \textit{i.e.} $\Omega_\mathrm{PBH}>\Omega_\mathrm{DM}$ . This tension can be alleviated by introducing a negative non-Gaussianity with $f_\mathrm{NL}\lesssim-1$ to suppress the PBH formation \cite{Franciolini:2023pbf,Liu:2023ymk,Choudhury:2023fwk}. Such a large and negative $f_\mathrm{NL}$ should be seen as the leading order of a fully nonlinear curvature perturbation, which can be realized in the curvaton scenario when the curvaton dominates at its decay moment \cite{Pi:2021dft,Ferrante:2022mui,Chen:2023lou}.
\item
It was shown that PBHs obey a Poisson distribution when the density perturbation is Gaussian \cite{Ali-Haimoud:2018dau,Desjacques:2018wuu}, while non-Gaussianity can make PBHs cluster initially on small scales \cite{Tada:2015noa,Young:2015kda,Franciolini:2018vbk,Young:2019gfc,Suyama:2019cst,DeLuca:2021hcf}, which enhances the merger rate of PBHs significantly \cite{Clesse:2016vqa,Raidal:2017mfl}. Such clustered PBHs can escape from the microlensing detections, which assume Poisson distribution \cite{Garcia-Bellido:2017xvr}. 

\item 
When the power spectrum of the curvature perturbation is enhanced on small scales, the loop corrections on CMB scales might be large, which may jeopardize the mechanism of enhancing the power spectrum on small scales to generate PBHs \cite{Kristiano:2022maq,Kristiano:2023scm}. It was soon realized that the loop correction can be suppressed when the transition to slow-roll inflation is smooth \cite{Riotto:2023gpm,Firouzjahi:2023aum,Firouzjahi:2023ahg,Firouzjahi:2023bkt}. 
For recent discussions, see \cite{Choudhury:2023rks,Motohashi:2023syh,Tasinato:2023ukp,Fumagalli:2023hpa,Cheng:2023ikq,Motohashi:2023syh,Franciolini:2023lgy,Fumagalli:2023loc,Maity:2023qzw,Tada:2023rgp,Iacconi:2023ggt,Davies:2023hhn,Firouzjahi:2024psd,Inomata:2024lud,Ballesteros:2024zdp,Kristiano:2024vst}.
\end{itemize}
For more discussion on the observational implications, see Part V (of the book \textit{Primordial Black Holes}) and the references therein, for instance the review papers \cite{Carr:2020gox,Carr:2020xqk,Carr:2023tpt}.

This paper is organized as follows. In Section \ref{sec8:routine} we review the calculation of PBH mass function, emphasizing on when and how the non-Gaussianity comes into play. In Section \ref{sec8:PNG}, we study some typical inflation models which can generate large non-Gaussianity: the ultra-slow-roll inflation, constant-roll inflation with a bumpy potential, single-field inflation with piecewise quadratic potential, and the curvaton scenario. After getting the non-Gaussian curvature perturbations in these models, we calculate the PBH mass function and abundance by the Press-Schechter-type formalism in Section \ref{sec8:abundance}, as well as the scalar-induced GWs in Section \ref{sec8:IGW}. We conclude by discussing some interesting topics to be studied in the future in Section \ref{sec8:conclusion}.

\section{A brief review of PBH formation}
\label{sec8:routine}

\subsection{PBH abundance in the Gaussian case}
In this section we will  briefly review how to calculate the PBH mass function and abundance by the Press-Schechter formalism, emphasizing on where and how the non-linearities and primordial non-Gaussianities enter. Our starting point is the power spectrum of the primordial curvature perturbation on comoving slices, $\mathcal{P_R}(k)$, calculated at the end of inflation. In the simplest Press-Schechter formalism, the density contrast on comoving slices is used, which is related to $\mathcal{R}$ by the Poisson equation
\begin{equation}\label{eqn8:delta-R}
\delta\equiv\left(\frac{\delta\rho}{\rho_0}\right)_c\approx-\frac{2(1+w)}{5+3w}\frac{1}{H^2a^2}\nabla^2\mathcal{R}.
\end{equation}
In radiation dominated case ($w=1/3$), the prefactor is 4/9. 

The density contrast threshold $\delta_\mathrm{th}$ is to determine whether a PBH could form in a specific Hubble patch or not, which was argued by the Jeans instability to be approximately $\delta_\text{th}\approx w$~\cite{Carr:1975qj}. A recent analytical result is the Harada-Yoo-Kohri (HYK) threshold \cite{Harada:2013epa}:
\begin{equation}\label{eqn8:HYK}
\delta_\text{th}=\frac{3(1+w)}{5+3w}\sin^2\left(\frac{\pi\sqrt w}{1+3w}\right)\xrightarrow{\quad w\to1/3\quad}0.41.
\end{equation}
Given the PDF of the curvature perturbation $\mathbb{P}(\mathcal{R})$, the PDF of the density contrast is given by
\begin{equation}\label{eqn8:PDFchain}
\mathbb{P}(\delta)=\mathbb{P}(\mathcal{R})\left|\frac{\mathrm{d}\mathcal{R}}{\mathrm{d}\delta}\right|,
\end{equation}
where the Jacobian $|\mathrm{d}\mathcal{R}/\mathrm{d}\delta|$ can be easily calculated from the linear Poisson equation \eqref{eqn8:delta-R}. With these quantities, we can calculate the PBH abundance in the following way:
\begin{equation}\label{eqn8:simplePS}
\left.
\begin{matrix}
\mathcal{R}\xrightarrow{\mathrm{Eq.~}\eqref{eqn8:delta-R}}\delta\\
\\
\mathbb{P}(\mathcal{R})\xrightarrow{\mathrm{Eq.~}\eqref{eqn8:PDFchain}}\mathbb{P}(\delta)
\end{matrix}
\right\}
\xrightarrow[\text{Window function}]{\delta_\mathrm{th}=0.41\mathrm{~(HYK~limit)}}
\beta=\int_{\delta_\mathrm{th}}\mathbb{P}(\delta)\frac{M(\delta)}{M_H}{\mathop{}\!\mathrm{d}}\delta.
\end{equation}
We will not discuss the issues of window function in this paper. The key point of the above calculation is that, if the curvature perturbation $\mathcal{R}$ is Gaussian, the linear approximation \eqref{eqn8:delta-R} tells us the density contrast is also Gaussian. Therefore the integration in \eqref{eqn8:simplePS} is simply a complementary error function, which gives a Gaussian suppression $\sim\exp(-\delta_\mathrm{th}^2/2\sigma_\delta^2)$ in the far tail.  
This simple estimation is frequently quoted in the literature. However, it depends crucially on the Gaussianity of $\mathcal{R}$ and the linear approximation \eqref{eqn8:delta-R}, both of which are doubtful when $\delta>\delta_\mathrm{th}$.

\subsection{Press-Schechter-type formalism}\label{sec8:EPS}

The estimation \eqref{eqn8:simplePS} neglects some important facts in the calculation of PBH abundance, which might change the result significantly.

\subsubsection{Compaction function}\label{sec8:compaction}
The criterion of PBH formation is not simply a threshold of the density contrast $\delta\rho/\rho$. Numerical studies on the black hole collapse show that the criterion should be put on the averaged mass excess inside a sphere of a fixed areal radius, dubbed the compaction function \cite{Shibata:1999zs}. 
In a reduced-circumference polar coordinates, we have
\footnote{In this paper, following Refs. \cite{Germani:2019zez,Escriva:2022duf,Germani:2023ojx}, we write the spatial metric as
\begin{align}\label{pertFRW}
\mathrm{d}\Sigma^2=e^{2\mathcal{R}(r)}\left(\mathrm{d}r^2+r^2\mathrm{d}\Omega^2\right)=\frac{\mathrm{d}\bar{r}^2}{1-K(\bar{r})\bar{r}^2}+\bar{r}^2\mathrm{d}\Omega^2
\end{align}
where $\mathrm{d}\Omega^2=\mathrm{d}\theta^2+\sin^2\theta\mathrm{d}\phi^2$ is the two-dimensional line element on a sphere. $\bar{r}$ is called reduced-circumference radius, which is related to $r$ by the local expansion:
\begin{align}
\bar{r}=e^{\mathcal{R}(r)}r.
\end{align}
}
\begin{equation}\label{eqn8:defC}
\mathscr{C}(t,\bar{r},\mathbf{x})\equiv\frac{2\delta M(t,\bar{r},\mathbf{x})}{\bar{r}}=\frac{3}{\bar{r}}\left(aH\right)^2\int^{\bar r}_0\mathrm{d\varrho}~\varrho^2\delta(t,\varrho,\mathbf{x}),
\end{equation}
where we set $G=1$. We can see that the compaction function is the averaged density contrast inside a sphere of radius $\bar{r}$, which is conserved on superhorizon scales. 
For the discussion of gauge issue, see \cite{Harada:2015yda}. 
It seems that, if we substitute the linear relation \eqref{eqn8:delta-R} into \eqref{eqn8:defC}, we have 
\begin{align}\label{eqn8:Cl}
\mathscr{C}_\ell=-\frac{6(1+w)}{5+3w}r\frac{\partial\mathcal{R}}{\partial r}\xrightarrow{\quad w=1/3\quad}-\frac{4}{3}r\frac{\partial\mathcal{R}}{\partial r}.
\end{align}
However, linear relation \eqref{eqn8:delta-R} is only the leading-order approximation of \cite{Harada:2015yda,Kawasaki:2019mbl,Young:2019yug,DeLuca:2019qsy} 
\begin{equation}\label{eqn8:nonlineardelta-R}
\delta=-\frac{4}{3}\cdot\frac{3(1+w)}{5+3w}\left(\frac{1}{aH}\right)^2e^{-5\mathcal{R}/2}\nabla^2e^{\mathcal{R}/2},
\end{equation}
Therefore we should use
\begin{align}\label{eqn8:Cnl}
\mathscr{C}=\mathscr{C}_{\ell}\left(1-\frac{5+3w}{12(1+w)}\mathscr{C}_{\ell}\right)\xrightarrow{\quad w=1/3\quad}
\mathscr{C}=\mathscr{C}_{\ell}-\frac{3}{8 } \mathscr{C}_{\ell}^2.
\end{align}
Such a nonlinear relation will naturally bring non-Gaussianity to the calculation of PBH formation even if $\mathcal{R}$ is Gaussian. 
From now on, we will focus on radiation-dominated universe with $w=1/3$.

\subsubsection{Threshold}

Around a density peak, the compaction function $\mathscr{C}(\bar{r})$ is a function of the radius $\bar{r}$. Numerical calculations tell us that whether an overdensed region can collapse into a black hole depends on whether the maximum of the compaction function, $\mathscr{C}(\bar{r}_m)$, exceeds the critical value $\mathscr{C}_\mathrm{th}$, which varies from 0.41 to 0.67, depending on the density profile of the overdensed region \cite{Musco:2018rwt}. It was further found that only the width of the compaction function peak, parameterized by $q\equiv-\mathscr{C}''(\bar{r}_m)\bar{r}_m^2/(4\mathscr{C}(\bar{r}_m))$, is important \cite{Escriva:2019phb}. 
Large $q$ corresponds to a narrow $\mathscr{C}(\bar{r}_m)$, given by a top-hat density profile which drops rapidly around $\bar{r}_m$. Strong pressure gradients at $\bar{r}_m$ can resist the gravitational collapse, thus $\mathscr{C}_\mathrm{th}$ reaches its maximum $2/3$ in this limit.
On the contrary,  small $q$ corresponds to a broad $\mathscr{C}(\bar{r}_m)$, given by a density profile highly peaked at the center. As the pressure gradients can be neglected, it reduces to the HYK limit
$\mathscr{C}_\mathrm{th}\approx\delta_\mathrm{th}\approx0.4$. 
In principle the density profiles and their statistical properties should be input if we want to calculate PBH abundance appropriately. Some hypothetical profiles are given by fitting the numerical results \cite{Musco:2018rwt,Escriva:2019phb,Young:2019osy}. In the theory of peaks, the profiles can be derived by the multipole moments of the power spectrum, provided the curvature perturbation is Gaussian 
\cite{Yoo:2018kvb,Atal:2019cdz,Atal:2019erb,Yoo:2020dkz,Kitajima:2021fpq}. 

By using \eqref{eqn8:Cnl}, $\mathscr{C}'(r)=0$ gives $\mathscr{C}_{\ell}'(r)=0$ or $\mathscr{C}_{\ell}(r)=4/3$. The former condition determines the extreme of $\mathscr{C}(r)$ at $r_m$, where its second derivative is
\begin{align}
\mathscr{C}''(r_m)=\left(1-\frac{3}{4}\mathscr{C}_{\ell}(r_m)\right)\mathscr{C}_{\ell}''(r_m).
\end{align}
We see that depending on whether $\mathscr{C}_{\ell}<4/3$ or $\mathscr{C}_{\ell}>4/3$, the maximum $\mathscr{C}(r_m)$ corresponds to the maximum or minimum of $\mathscr{C}_\ell$. Only the former case, dubbed type I fluctuation \cite{Kopp:2010sh}, is consistent with the linear approximation. Type II fluctuations with $\mathscr{C}_{\ell}>4/3$ are usually rarer, which we do not consider in this paper. Given the profile-dependent threshold on $\mathscr{C}$, the corresponding threshold on the linear compaction function $\mathscr{C}_\ell$ can be solved by \eqref{eqn8:Cnl} 
\begin{align}
\mathscr{C}_{\ell,\mathrm{th}}=\frac{4}{3}
\left(1-\sqrt{1-\frac{3}{2}\mathscr{C}_\mathrm{th}}\right).
\end{align}
As two examples, the threshold of a Maxican-hat profile considered in Chapter 7 \cite{Young:2024jsu} is 
$\mathscr{C}_\mathrm{th}\approx1/2$, which gives $\mathscr{C}_{\ell,\mathrm{th}}\approx2/3$. In the theory of peaks, a monochromatic power spectrum gives $\mathscr{C}_{\mathrm{th}}\approx0.587$, thus $\mathscr{C}_{\ell,\mathrm{th}}\approx0.872$ \cite{Kitajima:2021fpq}, which we will use later. 

When the power spectrum of $\mathcal{R}$ is broad, for every scale we study, we should add a window function to smooth out the unwanted short-wavelength fluctuations. This is highly non-trivial and there are uncertainties from the choice of the window function, which will not be discussed in this paper. 

\subsubsection{Primordial non-Gaussianity}
The primordial non-Gaussianity of $\mathcal{R}$ (if any) should be taken into account when deriving the PDF of $\mathscr{C}$ from the PDF of $\mathcal{R}$, which is the main topic of this paper.

We already know from Section \ref{sec8:compaction} that the compaction function $\mathscr{C}$ is a quadratic function of $\mathscr{C}_\ell$. In the presence of primordial non-Gaussianity in $\mathcal{R}$, even $\mathscr{C}_\ell$ is non-Gaussian. As we will see, for local non-Gaussianities, $\mathcal{R}$ is a function of a Gaussian-distributed variable, $\mathcal{R}_\mathrm{g}$, which obeys the Gaussian PDF
\begin{equation}\label{eqn8:P(Rg)}
\mathbb{P}(\mathcal{R}_\mathrm{g})=\frac{1}{\sqrt{2\pi\sigma^2_{\mathcal{R}}}}\exp\left(-\frac{\mathcal{R}_g^2}{2\sigma_\mathcal{R}^2}\right),
\end{equation}
with
\begin{equation}\label{eqn8:sigmaR}
\sigma_\mathcal{R}^2=\int\frac{\mathrm{d}k}{k}\tilde W^2(k,r)\mathcal{P}_{\mathcal{R}\mathrm{g}}(k).
\end{equation}
Local-type non-Gaussianity gives
\begin{equation}\label{eqn8:R(Rg)}
\mathcal{R}=\mathscr{F}[\mathcal{R}_\mathrm{g}(r)],\qquad
\mathcal{R}_\mathrm{g}=\mathscr{G}[\mathcal{R}(r)],
\end{equation}
where $\mathscr{G}\equiv\mathscr{F}^{-1}$ with $\mathscr{F}(0)=0$ and $\mathscr{F}'(0)=1$, such that \eqref{eqn8:R(Rg)} can be expanded as $\mathscr{F}(\mathcal{R}_\mathrm{g})\approx\mathcal{R}_\mathrm{g}$ when $\mathcal{R}_\mathrm{g}$ is small. Concrete examples of such nonlinear relations \eqref{eqn8:R(Rg)} will be discussed in the next section. 
From \eqref{eqn8:P(Rg)}, the corresponding PDF of $\mathcal{R}$ is
\begin{equation}\label{eqn8:Jacobian}
\mathbb{P}(\mathcal{R})=\mathbb{P}_{\mathrm{G}}(\mathcal{R}_\mathrm{g})\cdot\left| \frac{\mathrm{d}\mathscr{F} }{\mathrm{d} \mathcal{R}_\mathrm{g}}\right|^{-1}
=\mathbb{P}_{\mathrm{G}}(\mathcal{R}_\mathrm{g})\cdot\left| \frac{\mathrm{d}\mathscr{G} }{\mathrm{d} \mathcal{R}}\right|.
\end{equation}
This brings non-Gaussianity to the linear compaction function by
\begin{equation}\label{eqn8:Cl-Gg}
\mathscr{C}_\ell=-\frac{4}{3}r\frac{\mathrm{d}\mathscr{F}}{\mathrm{d}\mathcal{R}_\mathrm{g}}\frac{\partial\mathcal{R}_\mathrm{g}(r)}{\partial r}.
\end{equation}

\subsection{PBH abundance with primordial non-Gaussianity}
According to the discussion above, the PBH abundance should be calculated as following
\cite{Biagetti:2021eep,Gow:2022jfb,Ferrante:2022mui}
\begin{equation}\label{eqn8:routine2}
\left.
\begin{matrix}
\mathcal{R}\xrightarrow{\mathrm{Eq.~}\eqref{eqn8:Cl}}\mathscr{C}_\ell\xrightarrow{\mathrm{Eq.~}\eqref{eqn8:Cnl}}\mathscr{C}(\mathscr{C}_\ell)\\
\\
\left(
\begin{matrix}
\mathrm{PNG~from}\\
\mathrm{Inflation} 
\end{matrix}
\right)
\xrightarrow{}\mathbb{P}(\mathcal{R})\xrightarrow{\mathrm{Eq.~}\eqref{eqn8:Jacobian}}\mathbb{P}(\mathscr{C}_\ell)
\end{matrix}
\right\}
\xrightarrow[\text{Window function}]{\mathscr{C}_{\mathrm{th}}\mathrm{(profile)}}
\beta=\int^{4/3}_{\mathscr{C}_{\ell,\mathrm{th}}}\mathbb{P}(\mathscr{C}_\ell)\frac{M(\mathscr{C}_\ell)}{M_H}{\mathop{}\!\mathrm{d}}\mathscr{C}_\ell,
\end{equation}
where PNG stands for primordial non-Gaussianity of the curvature perturbation generated during inflation, which will be discussed in details in the next section. Calculations following the routines in \eqref{eqn8:routine2} are called extended Press-Schechter or Press-Schechter-type formalism, which is in principle different from another frequently used method, \textit{the theory of peaks}
\cite{Green:2004wb,Yoo:2018kvb,Germani:2018jgr,Atal:2019cdz,Atal:2019erb,Young:2020xmk,Yoo:2020dkz,Taoso:2021uvl,Riccardi:2021rlf,Kitajima:2021fpq,Young:2022phe},  by two main points. In Press-Schechter-type formalism, the PBH abundance is calculated by integrating the PDF of $\mathscr{C}_\ell$ from the threshold given by $\mathscr{C}_\ell\geq\mathscr{C}_{\ell,\mathrm{th}}$, while in the theory of peaks the PBH abundance is given by counting the number of peaks $n_\mathrm{pk}$ where the averaged compaction function in a sphere of radius $r_m$, $\langle\mathscr{C}\rangle_{r\leq r_m}$, exceeds a universal threshold 2/5 \cite{Escriva:2019phb}.

\section{Primordial non-Gaussianities of the curvature perturbation}\label{sec8:PNG}

Usually, the primordial curvature perturbation on comoving slices $\mathcal{R}$ originates from the quantum fluctuations of the inflaton $\varphi$ during inflation \cite{Brout:1977ix,Guth:1980zm,Starobinsky:1980te,Mukhanov:1981xt,Linde:1981mu,Albrecht:1982wi}. In linear perturbation theory,
$\mathcal{R}_g=-(H/\dot\varphi)\delta\varphi$ \cite{Mukhanov:1985rz,Sasaki:1986hm}, where $\delta\varphi$ is the field perturbation on spatially-flat slices. The non-Gaussian curvature perturbation is usually parameterized as the perturbative series of \eqref{eqn8:R(Rg)} \cite{Komatsu:2001rj},
\begin{equation}\label{eqn8:dN0}
\mathcal{R}=\mathcal{R}_g+\frac{3}{5}f_\mathrm{NL}\left(\mathcal{R}_g^2-\left\langle\mathcal{R}_g^2\right\rangle\right)+\cdots.
\end{equation}
In slow-roll inflation, the nonlinear parameter $f_\mathrm{NL}$ is related to the spectral tilt by the consistency relation $f_\mathrm{NL}=(5/12)(1-n_s)$~\cite{Maldacena:2002vr,Creminelli:2004yq}, which is suppressed by slow-roll parameters as $n_s-1=-2\epsilon_H-\eta_H$, with $\epsilon_H\equiv-\dot H/H^2$ and $\eta_H\equiv\dot\epsilon_H/(H\epsilon_H)$. On scales larger than 1 Mpc, the observational constraint is $f_\mathrm{NL}=-0.9\pm5.1$~\cite{Planck:2018jri,DES:2021wwk}. 

 As is commented above, the PBH abundance depends sensitively on the non-Gaussianities, thus even a small non-Gaussianity can change the PBH abundance significantly. Furthermore, as the power spectrum is enhanced rapidly (usually as power-law, $k^{\mathcal{O}(1)}$) before the peak scale, an intuitive guess based on the consistency relation $f_\mathrm{NL}\sim 1-n_s$ is that non-Gaussianity will also be enhanced to $\mathcal{O}(1)$. For instance, in ultra-slow-roll inflation, $\epsilon_H\lll1$ and $\eta_H=-6$. If we only focus on the slow-roll attractor mode, we have
\begin{align}\label{eqn8:fnl=5/2}
f_\mathrm{NL}=\frac{5}{12}(1-n_s)=\frac{5}{12}(2\epsilon_H+\eta_H)\approx\frac{5}{12}\times(-6)=-\frac{5}{2}.
\end{align}
Accidentally, this rough estimate gives the correct absolute value with an opposite sign \cite{Namjoo:2012aa}. Anyway this indicates that the non-Gaussianity may reach $\mathcal{O}(1)$ when the power spectrum increases rapidly.




In principle, when $|f_\mathrm{NL}|\gtrsim1$, the higher order terms include an infinite number of parameters, \textit{i.e.} $g_\mathrm{NL}$, $h_\mathrm{NL}$, $i_\mathrm{NL}$, \textit{etc}., which makes it impossible to have a model-independent analysis. 
Fortunately, in some typical models, the non-Gaussian curvature perturbation can be calculated neatly by $\delta N$ formalism, based on the \textit{separate universe} approach \cite{Sasaki:1995aw,Wands:2000dp,Lyth:2004gb}. During inflation, the gradient terms decay rapidly on superhorizon scales, which guarantees that in a region slightly larger than the Hubble patch, the inflaton field takes a homogeneous value, which varies from region to region. The initial condition for $\varphi$ in the numerous separate universes obeys a Gaussian distribution, which evolves independently on superhorizon scales. $\delta N$ formalism tells us that the final curvature perturbation in such a patch is the difference of total local expansion, \textit{i.e.} the $e$-folding number $N$, between this patch to the fiducial $\langle N\rangle$, evaluated from an initial spatially-flat slice to the final comoving slice:
\begin{align}\label{eqn8:deltaN}
\mathcal{R}=\delta N\equiv N(\varphi,\dot\varphi)-\langle N(\langle\varphi\rangle,\langle\dot\varphi\rangle)\rangle.
\end{align}
This formula does not require a decomposition of a background and a perturbation, thus can be used for large perturbations. We will use it to study the fully nonlinear curvature perturbation in the following sections. In Chapter 9 \cite{Vennin:2024yzl}, \eqref{eqn8:deltaN} will be used to study the quantum diffusion.

\subsection{Ultra-slow-roll inflation on a flat potential}\label{sec8:usr}

\begin{figure}[htbp]
\begin{center}
\includegraphics[width=0.9\textwidth]{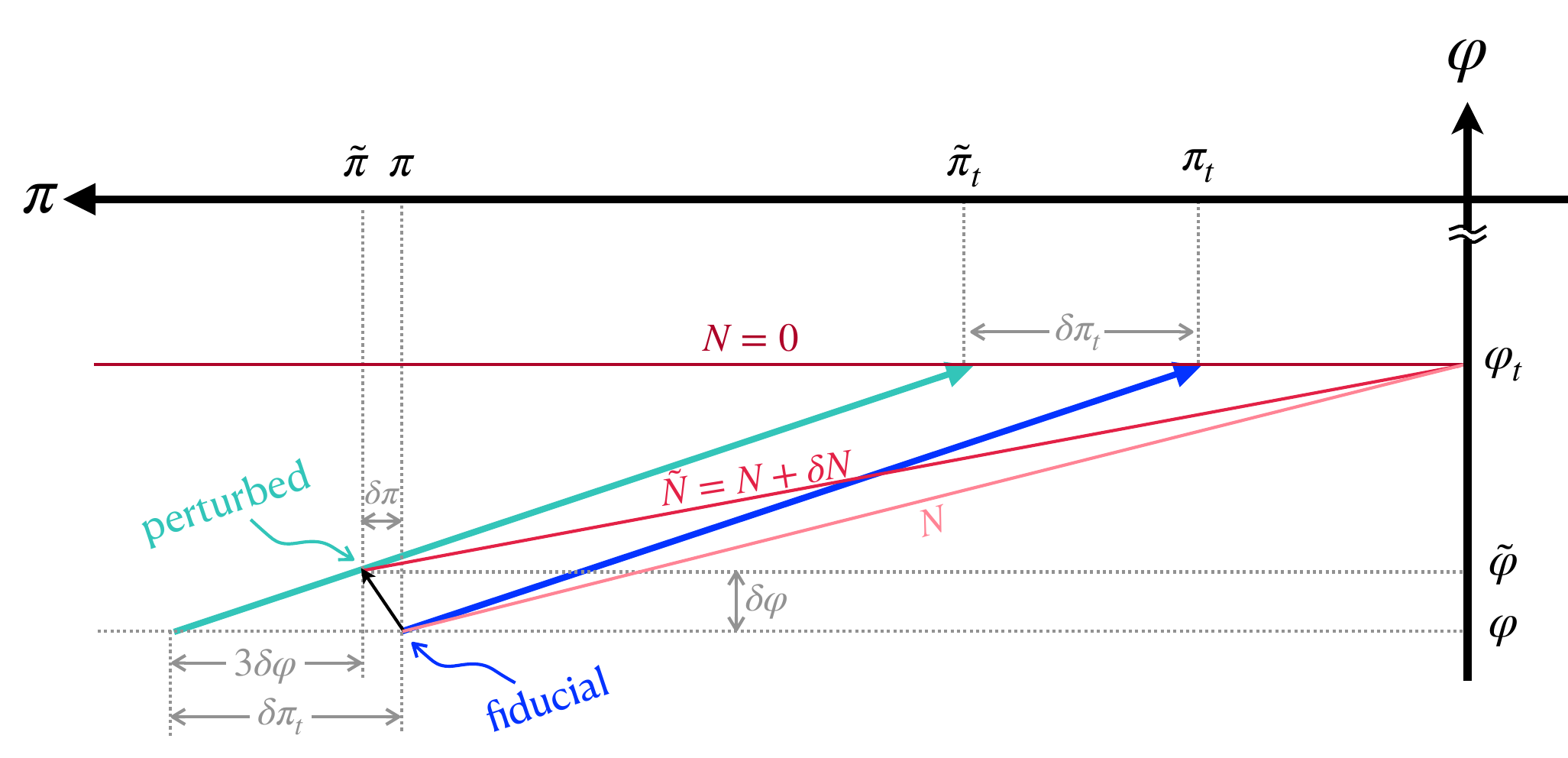}
\caption{The schematic phase portrait of the ultra-slow-roll inflation on a flat plateau. The blue and green parallel lines with arrows are the inflaton trajectories in the background universe and in a perturbed Hubble patch, respectively. The red lines are equal-$N$ lines, determined by the conservation law \eqref{eqn8:conservation-usr}, while ($\tilde\pi$, $\tilde\varphi$) at the perturbed trajectory corresponds to an $e$-folding number $\tilde{N}=N+\delta N$. Such a perturbed point ($\tilde\pi$, $\tilde\varphi$) satisfies \eqref{eqn8:conservation2-usr}, as is clearly shown geometrically. 
For more discussion on such a phase portrait, see \cite{Passaglia:2018ixg}.}
\label{fig8:usrphase}
\end{center}
\end{figure}

When $|f_\mathrm{NL}|\gtrsim1$, we should take into account all the higher order terms, which is only possible for some specific models. The simplest example is the so-called ultra-slow-roll inflation. Consider a plateau of potential $V(\varphi)=V_0$ for $\varphi_i<\varphi<\varphi_{t}$. On the plateau, Inflaton with an initial velocity moves toward the positive-$\varphi$ direction. The equation of motion of $\varphi$ is $\ddot\varphi+3H\dot\varphi=0$, where $H\approx\sqrt{V_0}/(\sqrt{3}M_\mathrm{Pl})$ is nearly a constant on the plateau ($M_\mathrm{Pl}=1/\sqrt{8\pi G}$). Conventionally, we define the $e$-folding number $N$, counted backward in time from $N(\varphi_{t})=0$, such that $\mathrm{d}N=-H\mathrm{d}t$. The equation of motion becomes
\begin{equation}\label{eqn8:eom-usr}
\frac{\mathrm{d}^2\varphi}{\mathrm{d}N^2}-3\frac{\mathrm{d}\varphi}{\mathrm{d}N}=0.
\end{equation}
Two general solutions to \eqref{eqn8:eom-usr} are $\varphi=\mathrm{constant}$ and $\varphi\propto e^{3N}$. The former static solution is the attractor, but it should not dominate at the classical level, otherwise the inflaton gets stuck on the plateau, which can only end inflation with the help of quantum diffusion. The non-attractor decelerating solution $\varphi\propto e^{3N}$ is necessary as it helps the inflaton to classically shoot out of the potential plateau with large enough initial velocity. We will see that the excitation of such non-attractor solutions is the origin of the large non-Gaussianity. Together with the boundary condition at the endpoint $\varphi_t$ of the plateau, the special solution is
\begin{align}\label{eqn8:N(phi)-usr}
\varphi(N)=\varphi_{t}+\frac{\pi_{t}}{3}\left(1-e^{3N}\right),
\end{align}
where $\pi_{t}$ is the velocity $\pi\equiv-\mathrm{d}\varphi/\mathrm{d}N$ at the endpoint where $N=0$, such that $\varphi(0)=\varphi_{t}$, $\pi(0)=\pi_{t}$. 
On the plateau, we have $\pi(N)=\pi_{t}e^{3N}$ as expected, and \eqref{eqn8:N(phi)-usr} can be rewritten as a ``conservation law"
\begin{equation}\label{eqn8:conservation-usr}
\pi(N)+3\varphi(N)=\pi_{t}+3\varphi_{t},
\end{equation}
which is valid for all the trajectories at any $e$-folding number. For a perturbed universe, \eqref{eqn8:conservation-usr} gives $\tilde\pi(\tilde N)+3\tilde\varphi(\tilde N)=\tilde\pi_{t}+3\varphi_{t}$, which in turn gives
\begin{equation}\label{eqn8:conservation2-usr}
\delta\pi+3\delta\varphi=\delta\pi_{t}.
\end{equation}
We defined $\delta\varphi\equiv\tilde\varphi-\varphi$, $\delta\pi\equiv\tilde\pi-\pi$ and $\delta\pi_{t}\equiv\tilde\pi_{t}-\pi_{t}$, which are Gaussian variables as $\varphi$ is a massless field. The physical meaning of \eqref{eqn8:conservation-usr} and \eqref{eqn8:conservation2-usr} can be well understood in the schematic phase portrait Fig. \ref{fig8:usrphase}.

From $\pi(N)=\pi_{t}e^{3N}$, we have
\begin{align}
N=\frac{1}{3}\ln\frac{\pi}{\pi_{t}}=-\frac{1}{3}\ln\left(1+\frac{3(\varphi-\varphi_{t})}{\pi}\right),
\end{align}
where the second equality comes from \eqref{eqn8:conservation-usr}, which is useful to determine the equal-$N$ lines in Fig.\ref{fig8:usrphase}: They are radial lines from $(\pi=0,~\varphi=\varphi_{t})$.

In the separate universe approach, another remote and independent ``perturbed universe'' happens to have another $e$-folding number of $\tilde N=(1/3)\ln \tilde\pi/\tilde\pi_{t}$, with $\tilde\pi=\pi+\delta\pi$, and $\tilde\pi_{t}=\pi_{t}+\delta\pi_{t}$. In the $\delta N$ formalism, the late-time curvature perturbation in this perturbed universe is the difference between these $e$-folding numbers
\begin{align}\label{eqn8:dN1}
\delta N=\tilde N-N=\frac{1}{3}\ln\left(\frac{\tilde\pi}{\tilde\pi_{t}}\frac{\pi_{t}}{\pi}\right)=\frac{1}{3}\ln\left(1-\frac{\delta\pi}{\pi}\right)-\frac{1}{3}\ln\left(1-\frac{\delta\pi_{t}}{\pi_{t}}\right)
\end{align}
Note that from from \eqref{eqn8:conservation2-usr} we have $\delta\pi\sim\delta\pi_{t}$, so unless $N\approx0$, the first term in \eqref{eqn8:dN1} is always neglegible compared to the second term as $\pi=\pi_{t}e^{3N}\gg\pi_{t}$.  If the later evolution do not contribute significantly to $\delta N$,\footnote{This requirement is not trivial. It is equivalent to require that the inflaton accelerates to the later slow-roll attractor after it classically shoots out of the flat pleateau. On the other hand, if $\varphi$ decelerates, the $\delta N$ contributed from the later stage is dominant, which is usually Gaussian and thus the entire curvature perturbation is nearly Gaussian. See Section \ref{sec8:logdual} for a detailed discussion.}
the final curvature perturbation is
\begin{equation}\label{eqn8:dN2}
\mathcal{R}\approx-\frac{1}{3}\ln\left(1-\frac{\delta\pi_{t}}{\pi_{t}}\right),
\end{equation}
 which is highly non-Gaussian with a nominal nonlinear parameter
\begin{equation}
f_\mathrm{NL}=\frac{5}{2}.
\end{equation}
Apparently, the non-Gaussianity expressed in \eqref{eqn8:dN2} can not be simply described by quadratic expansion with $f_\mathrm{NL}=5/2$.
We should emphasize that \eqref{eqn8:dN2} is only valid when the contribution from the later evolution to $\delta N$ is negligible, the meaning of which will be discussed in the subsection below.

\subsection{Constant-roll on a quadratic potential}\label{sec8:cr}
\begin{figure}[htbp]
\begin{center}
\includegraphics[width=0.48\textwidth]{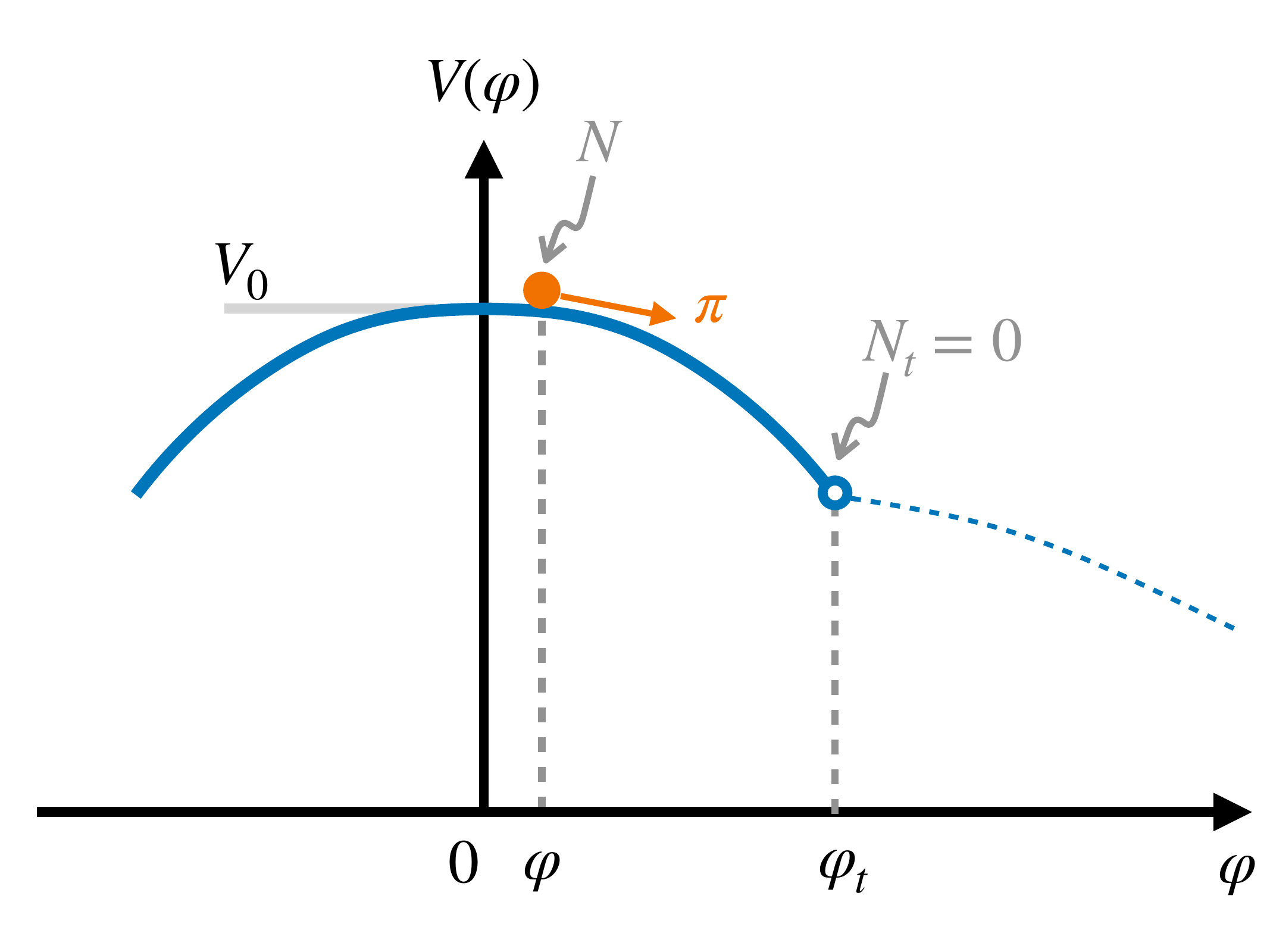}
\includegraphics[width=0.45\textwidth]{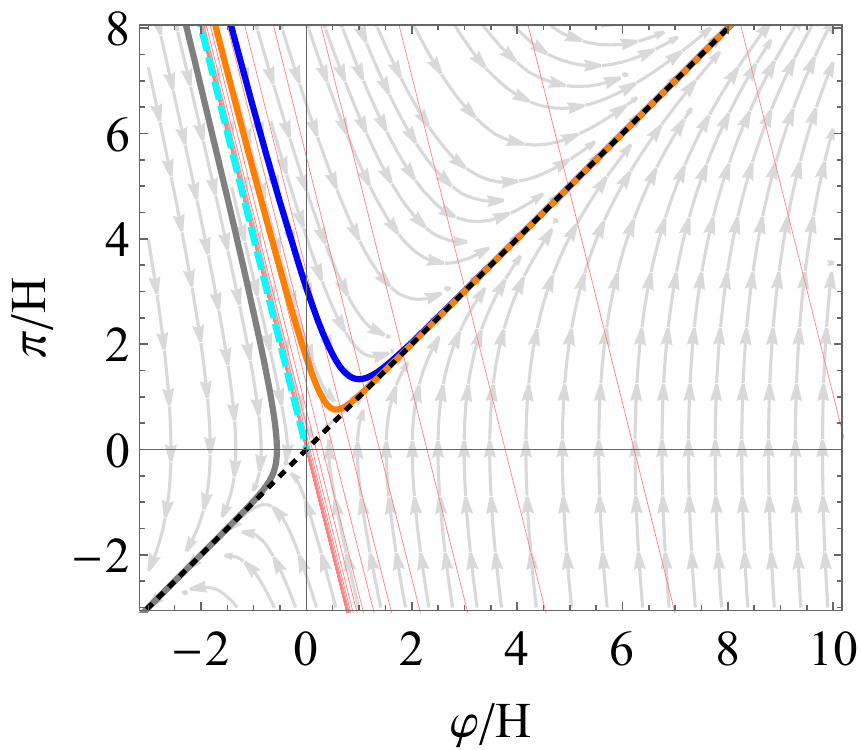}
\caption{Inflation potential \eqref{eqn8:bumpV}, and the phase portrait of \eqref{eqn8:eom} with $\eta=-4/3$. The initial conditions for the blue, orange, cyan, and gray curves are $\pi_i=14,\, 13,\, 12,\, 11$ at $\varphi_i=-3$ with $\varphi_{t}=100$ (in the unit of $H$). These solutions are not on the attractor $\pi+\lambda_-\varphi=0$ initially, which is displayed by the diagonal dotted black line. The equal-$N$ lines, based on \eqref{eqn8:N(phi)}, are shown in thin red lines with equal difference of $\Delta N=0.5$. It is clearly shown that for the same $\delta\varphi$, the largest $\delta N$ comes from around the critical trajectory (cyan dashed) in the constant-roll stage. }
\label{fig8:crphase}
\end{center}
\end{figure}

In the previous section we see how the non-attractor solution is important in enhancing the power spectrum and generating large non-Gaussianities. Such a conclusion is general, valid also for other types of potentials. In this subsection we will consider another analytically solvable model -- constant-roll inflation on a quadratic potential, which can be inserted as an intermediate stage between two slow-roll stages to enhance the curvature perturbation. We consider the following potential 
\begin{align}\label{eqn8:bumpV}
V(\varphi)=V_0+\frac{1}{2}m^2\varphi^2, 
\end{align}
shown in Fig.\ref{fig8:crphase}. We assume $V_0\gg |m^2\varphi^2|$, such that the Hubble parameter can still be approximated by a constant $H\approx \sqrt{V_0}/\sqrt3M_\mathrm{Pl}$, while the second slow-roll parameter $\eta\equiv m^2/3H^2$ is also a constant. The equation of motion is
\begin{equation}\label{eqn8:eom}
\frac{{\mathop{}\!\mathrm{d}}^2\varphi}{{\mathop{}\!\mathrm{d}} N^2}-3\frac{{\mathop{}\!\mathrm{d}}\varphi}{{\mathop{}\!\mathrm{d}} N}+3\eta\varphi=0,
\end{equation}
which is written in the $e$-folding number $N$, counted backwards in time from a junction point $\varphi_t$ far enough from the bump, such that the solution is already an attractor there \footnote{For the curvature perturbation which exits the horizon during the constant-roll stage, there will be still some $e$-folding number from $\varphi_t$ to the end of inflation, say $\varphi_e$. However, if the trajectories we study here already merged into the attractor solution before $\varphi_t$, these Hubble patches evolves uniquely after $\varphi_{t}$, thus the contribution to $\delta N$ from $\varphi_{t}$ to $\varphi_{e}$ is negligible.}.
Setting $\varphi\propto e^{\lambda N}$, the characteristic root $\lambda$ of \eqref{eqn8:eom} is
\begin{equation}\label{eqn8:lambdapm}
\lambda^2-3\lambda+3\eta=0,\quad\Longrightarrow\quad \lambda_\pm=\frac{3\pm\sqrt{9-12\eta}}{2}.
\end{equation}
For $\eta<3/4$, we have two real solutions with $\lambda_-<\lambda_+$, which means $\varphi$ has two independent solutions: the attractor solution $\varphi\propto e^{\lambda_-N}$ and the non-attractor solution $\varphi\propto e^{\lambda_+N}$. Again, defining the velocity $\pi(N)=-\mathrm{d}\varphi/\mathrm{d}N$, the attractor and non-attractor solutions satisfy
\begin{equation}
\pi+\lambda_\mp\varphi=0.
\end{equation}
Considering the boundary conditions $\pi(0)=\pi_t$, $\varphi(0)=\varphi_t$, the special solution is 
\begin{equation}\label{eqn9:vf1}
\varphi(N)=-\frac{\pi_t+\lambda_{-}\varphi_{t}}{\lambda_+-\lambda_-}e^{\lambda_+N}
+\frac{\pi_t+\lambda_{+}\varphi_{t}}{\lambda_+-\lambda_-}e^{\lambda_-N}.
\end{equation}
Therefore, we can solve these equations for $N$:
\begin{align}\label{eqn8:N(phi)}
N=\frac{1}{\lambda_\pm}\ln\frac{\pi+\lambda_\mp\varphi}{\pi_t+\lambda_\mp\varphi_{t}}\,,
\end{align}

The two different expressions for $N$ in \eqref{eqn8:N(phi)} are equivalent under the ``conservation law''
\begin{equation}\label{eqn8:conservation2}
\left(\frac{\pi+\lambda_+\varphi}{\pi_t+\lambda_+\varphi_{t}}\right)^{\lambda_+}
=\left(\frac{\pi+\lambda_-\varphi}{\pi_t+\lambda_-\varphi_{t}}\right)^{\lambda_-},
\end{equation}
where both sides equal to $e^{\lambda_+\lambda_-N}=e^{3\eta N}$, a direct consequence of the solution \eqref{eqn9:vf1}. For a perturbed universe where $\tilde\varphi=\varphi+\delta\varphi$, $\tilde\pi=\pi+\delta\pi$, $\tilde\pi_t=\pi_t+\delta\pi_t$, $\tilde N=N+\delta N$, we have
\begin{align}\label{eqn8:tildeN(phi)}
N+\delta N=\frac{1}{\lambda_\pm}\ln\frac{\pi+\delta\pi+\lambda_\mp\varphi}{\pi_t+\delta\pi_t+\lambda_\mp\varphi_t}\,,
\end{align}
Subtracting \eqref{eqn8:N(phi)} from \eqref{eqn8:tildeN(phi)}, we have
\begin{align}\label{eqn8:dN3}
\mathcal{R}=\delta N=\frac{1}{\lambda_\pm}\ln\left(1+\frac{\delta\pi+\lambda_\mp\delta\varphi}{\pi+\lambda_\mp\varphi}\right)
-\frac{1}{\lambda_\pm}\ln\left(1+\frac{\delta\pi_t}{\pi_t+\lambda_\mp\varphi_t}\right)\,.
\end{align}
We see that the curvature perturbation has two equivalent expressions, which is called \textit{logarithmic duality} in Ref. \cite{Pi:2022ysn}. Apparently it comes from the quadratic nature of the potential, which is a good approximation when the field excursion we study is not too long. The equivalence of the upper and lower formulas in \eqref{eqn8:dN3} can be easily shown by the perturbed version of conservation law \eqref{eqn8:conservation2}:
\begin{align}\label{eqn8:conservation3}
\left(1+\frac{\delta\pi_{t}}{\pi_{t}+\lambda_+\varphi_{t}}\right)^{-\lambda_+}\left(1+\frac{\delta\pi_{t}}{\pi_{t}+\lambda_-\varphi_{t}}\right)^{\lambda_-}=\left(1+\frac{\delta\pi+\lambda_-\delta\varphi}{\pi+\lambda_-\varphi}\right)^{\lambda_-}\left(1+\frac{\delta\pi+\lambda_+\delta\varphi}{\pi+\lambda_+\varphi}\right)^{-\lambda_+}\,.
\end{align}
In principle, the perturbation at the boundary $\delta\pi_{t}$ induced by the initial perturbation ($\delta\varphi$, $\delta\pi$) should be solved by \eqref{eqn8:conservation3}. However, in this case we do not need it after assuming that the inflaton is already the attractor solution at $\varphi_{t}$, \textit{i.e.} $\pi_{t}+\lambda_-\varphi_{t}\approx0$ and $\delta\pi_{t}\approx0$. Therefore, it is more convenient to use the lower line of \eqref{eqn8:dN3}, as the second logarithm is negligible:
\begin{align}\label{eqn8:dN4}
\mathcal{R}\approx\frac{1}{\lambda_-}\ln\left(1+\frac{\delta\pi+\lambda_+\delta\varphi}{\pi+\lambda_+\varphi}\right)
=\frac{1}{\lambda_-}\ln\left(1-\left.\lambda_-\frac{(\delta\varphi)_k}{\pi}\right|_{a\gg k/H}\right).
\end{align}
In the second step we use the fact that $\delta\varphi$ and $\varphi$ obeys the same equation of motion \eqref{eqn8:eom} on superhorizon scales (i.e. $k^2\ll H^2a^2$), when the potential is quadratic \footnote{This is a special case of the well know proportionality $\delta\varphi\propto\dot\varphi/H$ on superhorizon scales for \textit{any} potential in single-field inflation. }. Therefore, $(\delta\pi+\lambda_+\delta\varphi)/(\pi+\lambda_+\varphi)$ is a constant along each trajectory, which can be evaluated at a much later time after horizon exit (marked by $a\gg k/H$) when $\varphi$ is already in the attractor (but still before the boundary) \cite{Atal:2019cdz}. 
Note that $(\delta\varphi)_k/\pi$ in \eqref{eqn8:dN4} was stretched out of the horizon at $a_k=k/H$, which is now on a much larger scale of $a/k\gg1/H$. It can be calculated by evolving the solution of Mukhanov-Sasaki equation to late time $a\gg k/H$, which should not be confused with the ``new'' perturbations exiting the horizon at that moment. The phase diagram is shown in Fig.\ref{fig8:crphase}.

Slow-roll inflation and ultra-slow-roll inflation discussed in Section \ref{sec8:usr} are special cases of \eqref{eqn8:dN3}. Slow roll is the attractor solution when $\lambda_-\ll1$, $\lambda_+\approx3$. Expanding \eqref{eqn8:dN4}, we get \eqref{eqn8:dN0} with $\mathcal{R}_g=-\delta\varphi/\pi=H\delta\varphi/\dot\varphi$ and $f_\mathrm{NL}=-(5/6)\lambda_-$. In ultra-slow-roll inflation, the non-attractor solution dominates. Substituting $\lambda_-=0$, $\lambda_+=3$ into the upper line of Eq. \eqref{eqn8:dN3}, we can get \eqref{eqn8:dN1}.

\subsection{A general analysis on single-field inflation}\label{sec8:logdual}
We have shown how a quadratic piece of the potential \eqref{eqn8:bumpV} can give rise to the logarithmic relation \eqref{eqn8:dN3} of $\mathcal{R}(\delta\varphi,\delta\pi)$. The Gaussian variables $(\delta\varphi,\delta\pi)$ originate from quantum fluctuations on subhorizon scales, which should be evaluated on the initial spatially-flat surface after the horizon exit. In general, such a potential \eqref{eqn8:bumpV} is a segment of a global potential. In a typical three-segment model, the initial condition of $(\varphi, \pi)$ are usually provided by a preceding (slow-roll) stage, while after the bump/plateau, the potential is usually followed by another slow-roll potential to end inflation. Before and after the junction point, the attractor solutions are usually different. This can excite the non-attractor solution transiently, which contribute to the total $\delta N$ as we discussed above. All of such contributions should be taken into account to calculate the final $\delta N$.

\begin{figure}[htbp]
\begin{center}
\includegraphics[width=0.45\textwidth]{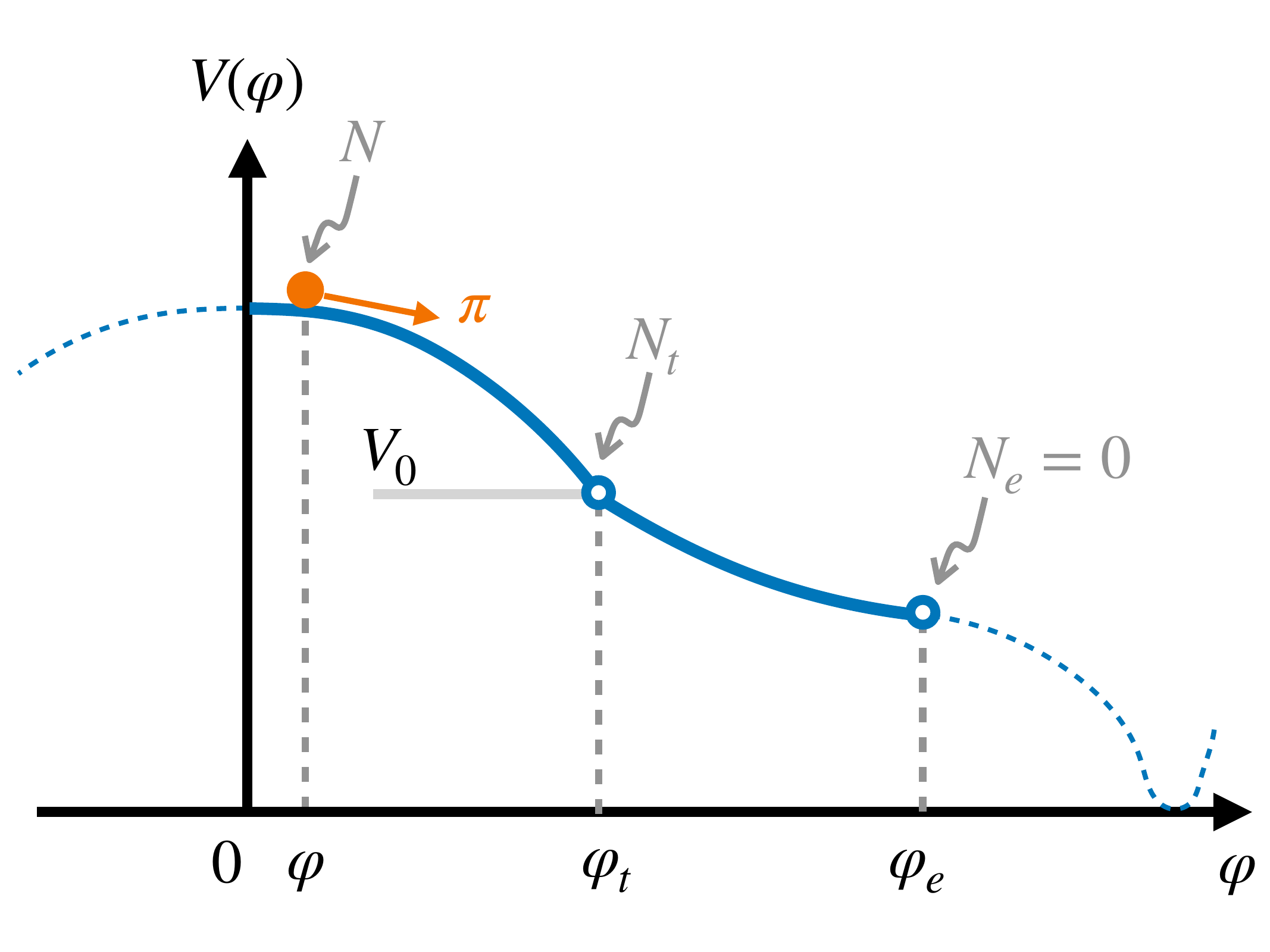}
\includegraphics[width=0.45\textwidth]{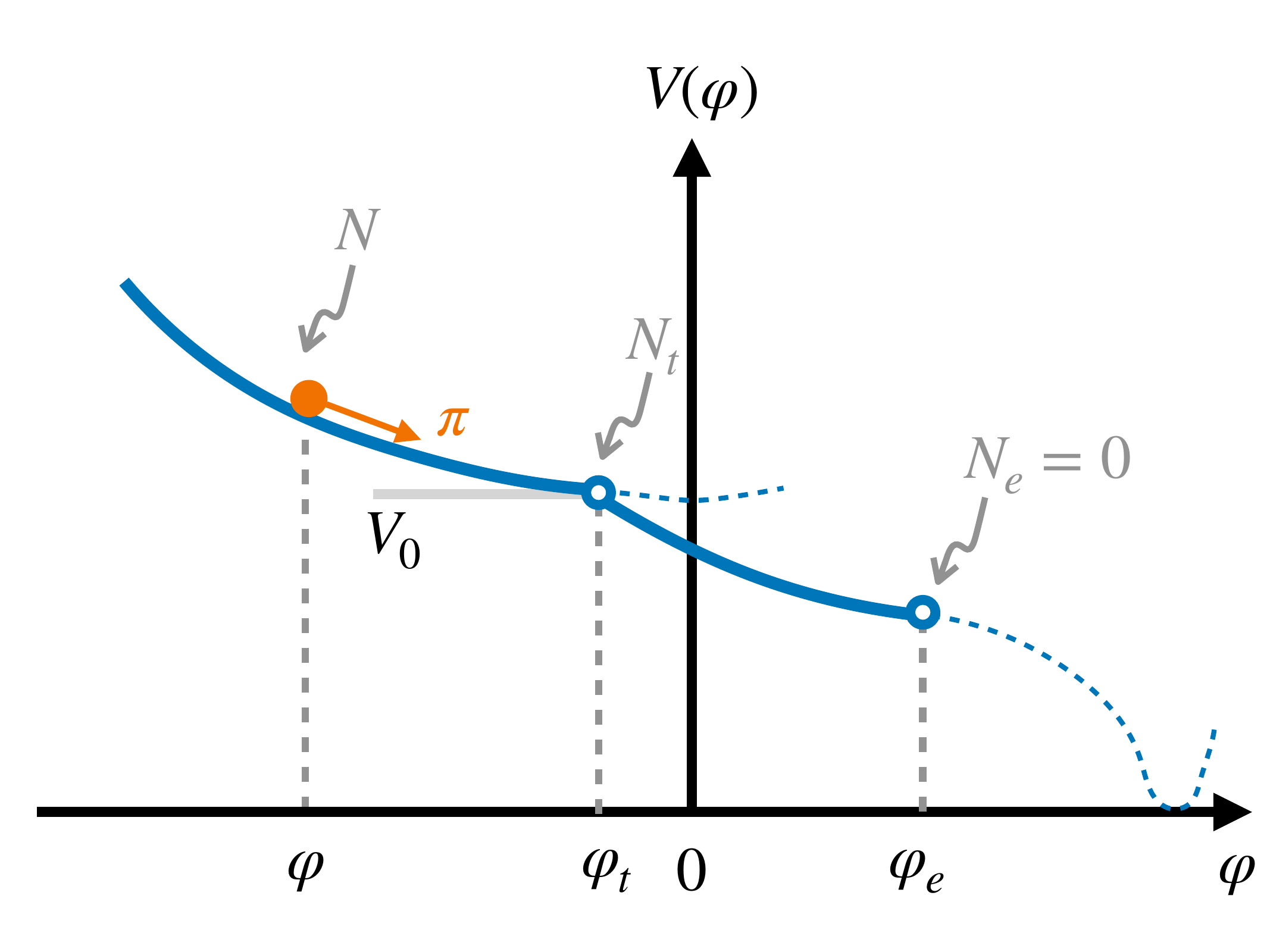}
\caption{Schematic pictures of the piecewise quadratic potential $V(\varphi)$ in \eqref{eqn8:piesewiseV}, with $m^2<0$ (left) and $m^2>0$ (right). The potential is continuous at the transition point $\varphi_{t}$, where $V(\varphi_{t})=V_0$, but $V'(\varphi_{t})$ has a jump. The origin is chosen at the local maximum (left panel) or minimum (right panel) of $V(\varphi<\varphi_t)$, for $m^2<0$ or $m^2>0$, respectively. The $e$-folding numbers are labeled at $\varphi$, $\varphi_{t}$, and $\varphi_{e}$.}
\label{fig8:potential}
\end{center}
\end{figure}

We consider a piecewise potential consisting of two parabolas:
\begin{align}\label{eqn8:piesewiseV}
V(\varphi)=\left\{
\begin{matrix}
&\displaystyle V_0+\frac{m^2}{2}\left(\varphi^2-\varphi_{t}^2\right), & \text{for~}\varphi\leq\varphi_{t}; \\
\\
&\displaystyle V_0-\frac{m'^2}{2}\left(\varphi_{t}-\varphi_m\right)^2+\frac{m'^2}{2}\left(\varphi-\varphi_m\right)^2,
& \text{for~}\varphi>\varphi_{t}.
\end{matrix}
\right.
\end{align}
where $\varphi_{t}$ is the junction point of the two quadratic potentials with $V(\varphi_{t})=V_0$, and  $\varphi_m$ is the minimum of $V(\varphi>\varphi_{t})$. For simplicity, we set the origin to make $\varphi=0$ the maximum (minimum) of $V(\varphi<\varphi_t)$, when $m^2<0$ ($m^2>0$). This is an extension of the Starobinsky model with a hinged linear potential \cite{Starobinsky:1992ts} to quadratic level, which can be further extended to a multi-exponential potential \cite{Domenech:2023dxx}. Inflation ends at $\varphi_{e}$ in the second stage. A schematic figure of this piecewise potential is shown in Fig.~\ref{fig8:potential}. $V(\varphi)$ is continuous, but $V'(\varphi)$ has a jump at $\varphi_{t}$. This means $\varphi'$ is continuous, but a change in the slope means that the attractor solution for $\varphi<\varphi_{t}$ is not the attractor solution for $\varphi>\varphi_{t}$, thus $\varphi$ will decelerate/accelerate to reach the later attractor solution, during which the curvature perturbation will be enhanced/suppressed, and large non-Gaussianity could be generated. 

By the method used in Section \ref{sec8:cr}, we can solve $\delta N$ contributed in all the stages, and add together to get the final curvature perturbation \cite{Pi:2022ysn}:
\begin{align}\label{eqn8:dN5}
\mathcal{R}&=\delta N_1\left(\{\varphi,\pi\}\to\{\varphi_{t},\pi_{t}\}\right)+\delta N_2\left(\{\varphi_{t},\pi_{t}\}\to\{\varphi_{e},\pi_{e}\}\right),\\\label{eqn8:dN6}
\delta N_1&=\frac{1}{\lambda_\pm}\ln\left(1+\frac{\delta\pi+\lambda_\mp\delta\varphi}{\pi+\lambda_\mp\varphi}\right)-\frac{1}{\lambda_\pm}\ln\left(1+\frac{\delta\pi_{t}}{\pi_{t}+\lambda_\mp\varphi_{t}}\right),\\\label{eqn8:dN7}
\delta N_2&=\frac{1}{\lambda'_\pm}\ln\left(1+\frac{\delta\pi_{t}}{\pi_{t}+\lambda'_\mp(\varphi_{t}-\varphi_m)}\right)
-\frac{1}{\lambda'_{\pm}}\ln\left(1+\frac{\delta\pi_{e}}{\pi_{e}+\lambda'_\mp(\varphi_{e}-\varphi_m)}\right)\,,
\end{align}
where $\lambda_\pm$ and $\lambda'_\pm$ are the characteristic roots of each piece, defined by \eqref{eqn8:lambdapm}. The equivalence of upper and lower expressions of each $\delta N_i$ are guaranteed by the perturbed conservation law \eqref{eqn8:conservation3} for each piece. Here, $\delta\varphi$ and $\delta\pi$ are the inflaton perturbation and its momentum (\textit{i.e.} time derivative with respect to $-N$) on an initial spatially-flat slice around ($\varphi$, $\pi$), which should be solved from the equation of motion for $\delta\varphi$, and evaluated at a moment slightly later than the horizon exit. In slow-roll stage, typically $\delta\varphi\approx H/(2\pi)$ and $\delta\pi\approx0$ as $\delta\pi$ decays exponentially after horizon exit, yet it is not so simple for non-attractor stages. For a concrete example of calculating $\delta\varphi$ and $\delta\pi$ in ultra-slow-roll inflation, see \cite{Jackson:2023obv,Domenech:2023dxx}. After we know $\delta\varphi$ and $\delta\pi$, the corresponding perturbations on all the later boundaries, \textit{i.e.} $\delta\pi_{t}$, $\delta\pi_{e}$, \textit{etc}., can be solved by the perturbed conservation law \eqref{eqn8:conservation3} for each piece, although this equation can only be solved numerically except for a few special values of $\lambda_-$ listed in Ref. \cite{Pi:2022ysn}. As we have shown before, finally all the different trajectories induced by the initial perturbation ($\delta\pi$, $\delta\varphi$) merge to the unique attractor trajectory. Since then, the $e$-folding number only depends on $\varphi$, and $\delta N$ in this stage only depends on the time difference when $\varphi$ falls on to the attractor, which is usually Gaussian. It is straightforward to extend \eqref{eqn8:dN5} to include more segments of quadratic potentials, and for every segment of the quadratic piece, two logarithms from the early and late boundaries are present, similar to \eqref{eqn8:dN7}. 


\begin{figure}[htbp]
\begin{center}
\includegraphics[width=0.495\textwidth]{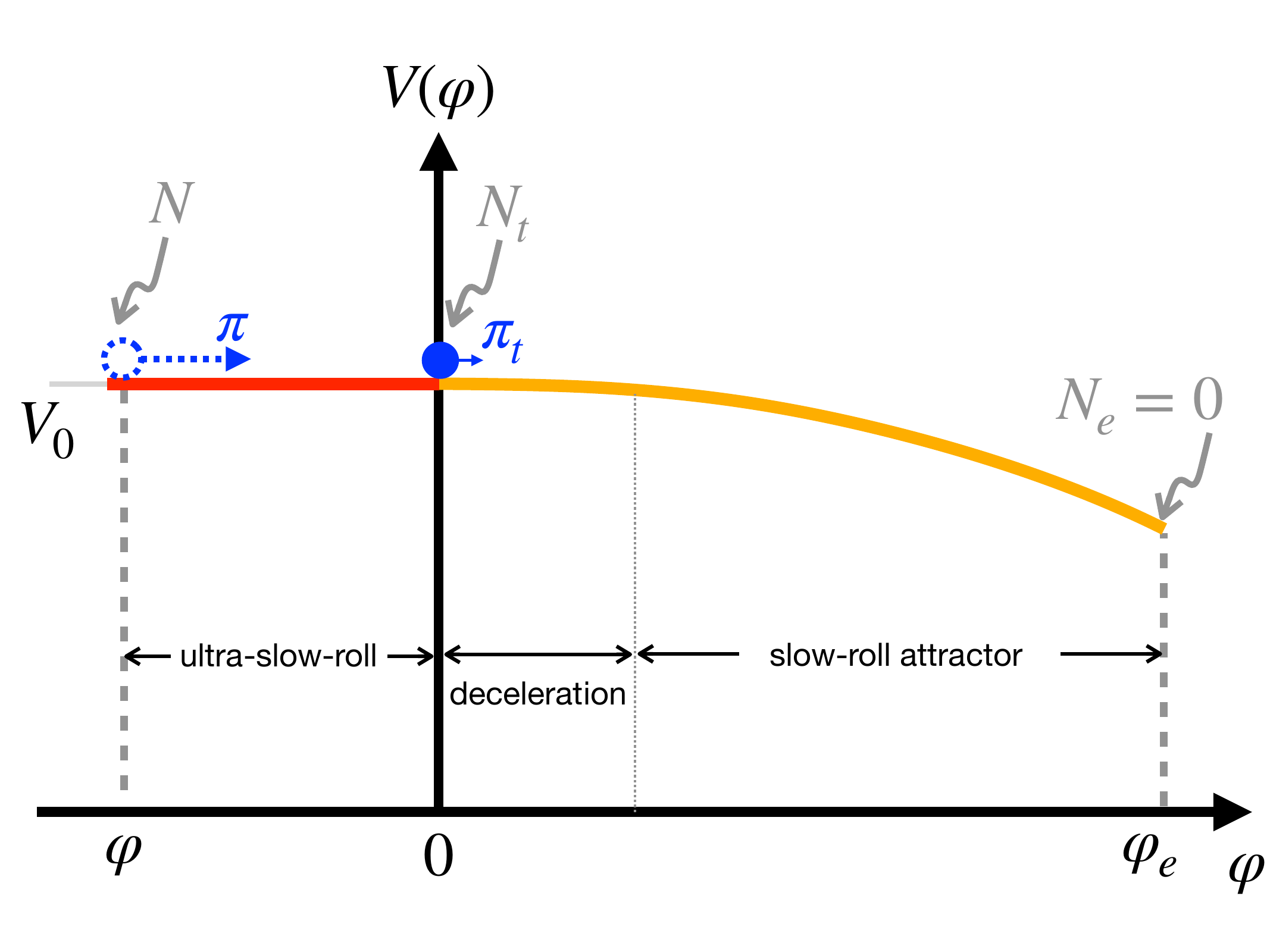}
\includegraphics[width=0.495\textwidth]{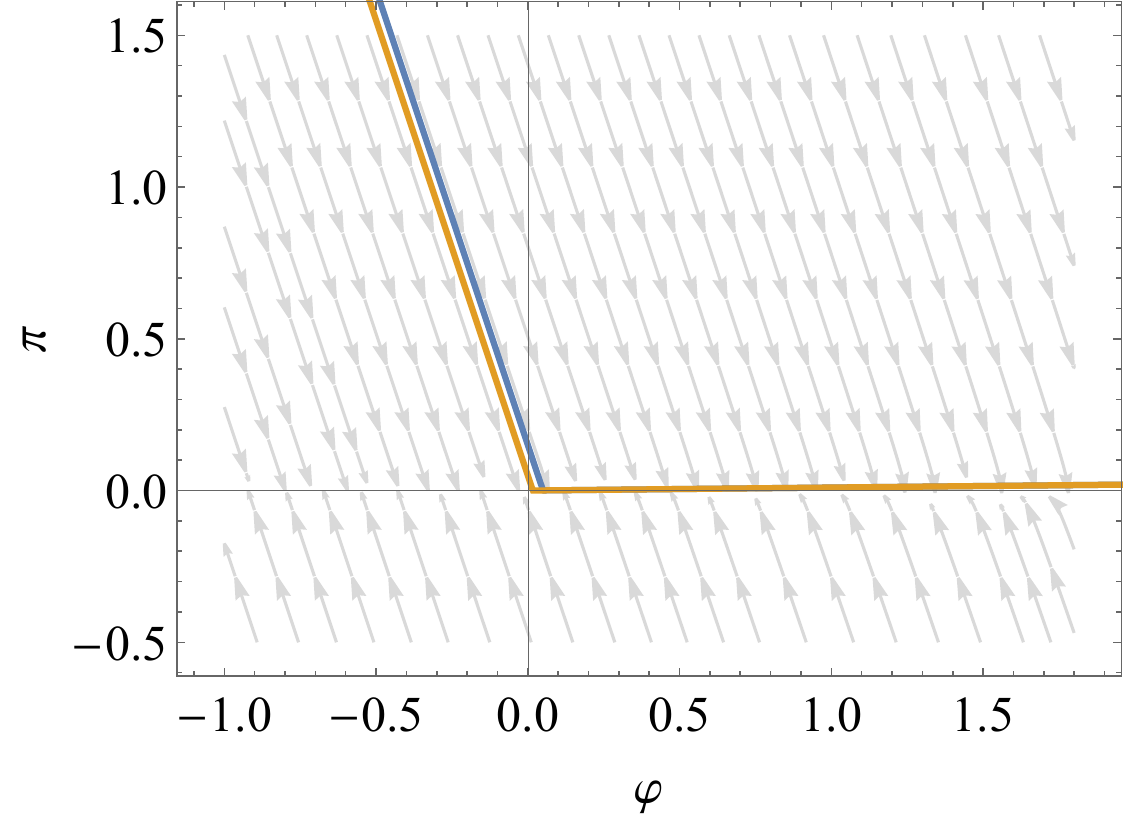}
\caption{The schematic potential (left panel) and the phase portrait (right panel) of the ``smooth transition'' from ultra-slow-roll to slow-roll, with $\eta'=-0.01$, $\varphi_m=0$. For fiducial (blue) and perturbed (orange) trajectories we set $\pi=9.15,~9.05$ at $\varphi=-3$ (in the unit of $H$), respectively. To show the difference of the trajectories clearly, $\delta\pi$ is exaggerated. Equal-$N$ lines are parallel to the fast-roll streams (\textit{i.e.} $\pi\propto-3\varphi$) with almost equal intervals, which are not drawn for visual clearance. This clearly shows that the contribution to $\delta N$ from the ultra-slow-roll stage is negligible, and the contribution from the later stage preserves slow-roll-type Gaussianity.}
\label{fig8:phase2}
\end{center}
\end{figure}

As a simple example, let's discuss ultra-slow-roll inflation on a flat potential connected to a slow-roll potential. Then \eqref{eqn8:dN6} reduces to \eqref{eqn8:dN2}, while \eqref{eqn8:dN7} reduces to \eqref{eqn8:dN4}:
\begin{align}\label{eqn8:dN8}
\mathcal{R}\approx-\frac{1}{3}\ln\left(1+\frac{\delta\pi_{t}}{\pi_{t}}\right)+\frac{1}{\lambda'_-}\ln\left(1+\frac{\delta\pi_{t}}{\pi_{t}+\lambda'_+\left(\varphi_{t}-\varphi_{m}\right)}\right).
\end{align}
In general, the final curvature perturbation is a sum of these two terms, and the PDF of $\mathcal{R}$ can only be solved numerically when two logarithmic functions are equally important. However, it can be solved analytically when one of the logarithms dominates. When $\pi_{t}\gg\lambda_-'(\varphi_{t}-\varphi_{m})$, the velocity at the end of flat plateau surpasses the attractor velocity of the slow-roll potential, thus a transient deceleration stage is followed. The second term of \eqref{eqn8:dN8} dominates, so the final curvature perturbation is nearly Gaussian with $f_\mathrm{NL}=-(6/5)\lambda_-'\approx-(6/5)\eta'$. On the contrary, when $\pi_{t}\ll\lambda_-'\left(\varphi_{t}-\varphi_{m}\right)$, the velocity at the end of flat plateau is not yet the attractor of the slow-roll potential, thus a transient acceleration stage is followed. The first term of \eqref{eqn8:dN8} dominates, so the final curvature perturbation is non-Gaussian with $f_\mathrm{NL}=5/2$. These two limits are called ``smooth transition'' and ``sharp transition'' respectively in \cite{Cai:2018dkf}, where a parameter $h\equiv6|\lambda'_-(\varphi_{t}-\varphi_{m})/\pi_{t}|$ was defined to compare the slow-roll attractor velocity $-\lambda'_-(\varphi_{t}-\varphi_{m})$ and the realistic end-of-ultra-slow-roll velocity $\pi_{t}$ at the transition point $\varphi_{t}$. To avoid confusion, we should emphasize that the comparison is between the velocities, but not between the slopes of the potentials.

This result is easy to understand by looking at the phase portrait. For the ``smooth transition'' case, although the potential behaves differently for $\varphi<\varphi_{t}$ and $\varphi>\varphi_{t}$, but the non-attractor solution ($\pi=-3\varphi$) in the flat plateau is almost the same as the non-attractor solution in a slow-roll potential ($\pi=-\lambda_+'\varphi\approx(3-\eta')\varphi$). Therefore, this case is similar to the bumpy potential we studied in Section \ref{sec8:cr}, and \eqref{eqn8:dN4} tells us $\mathcal{R}$ is nearly Gaussian with $f_\mathrm{NL}=-(6/5)\lambda_-'$. The phase portrait Fig. \ref{fig8:phase2} looks similar to that of a slow-roll inflation, as the ultra-slow-roll trajectories are similar to the fast-roll trajectories in an ordinary slow-roll inflation, which are parallel to the equal-$N$ lines. The difference in the $e$-folding number for two nearby trajectories are proportional to the difference in $\varphi$, therefore $\delta N\propto\delta\varphi$ which preserves Gaussianity. The well studied Starobinsky model \cite{Starobinsky:1992ts,Leach:2001zf,Biagetti:2018pjj,Tasinato:2020vdk,Pi:2022zxs,Domenech:2023dxx} belongs to this limit.

\begin{figure}[htbp]
\begin{center}
\includegraphics[width=0.495\textwidth]{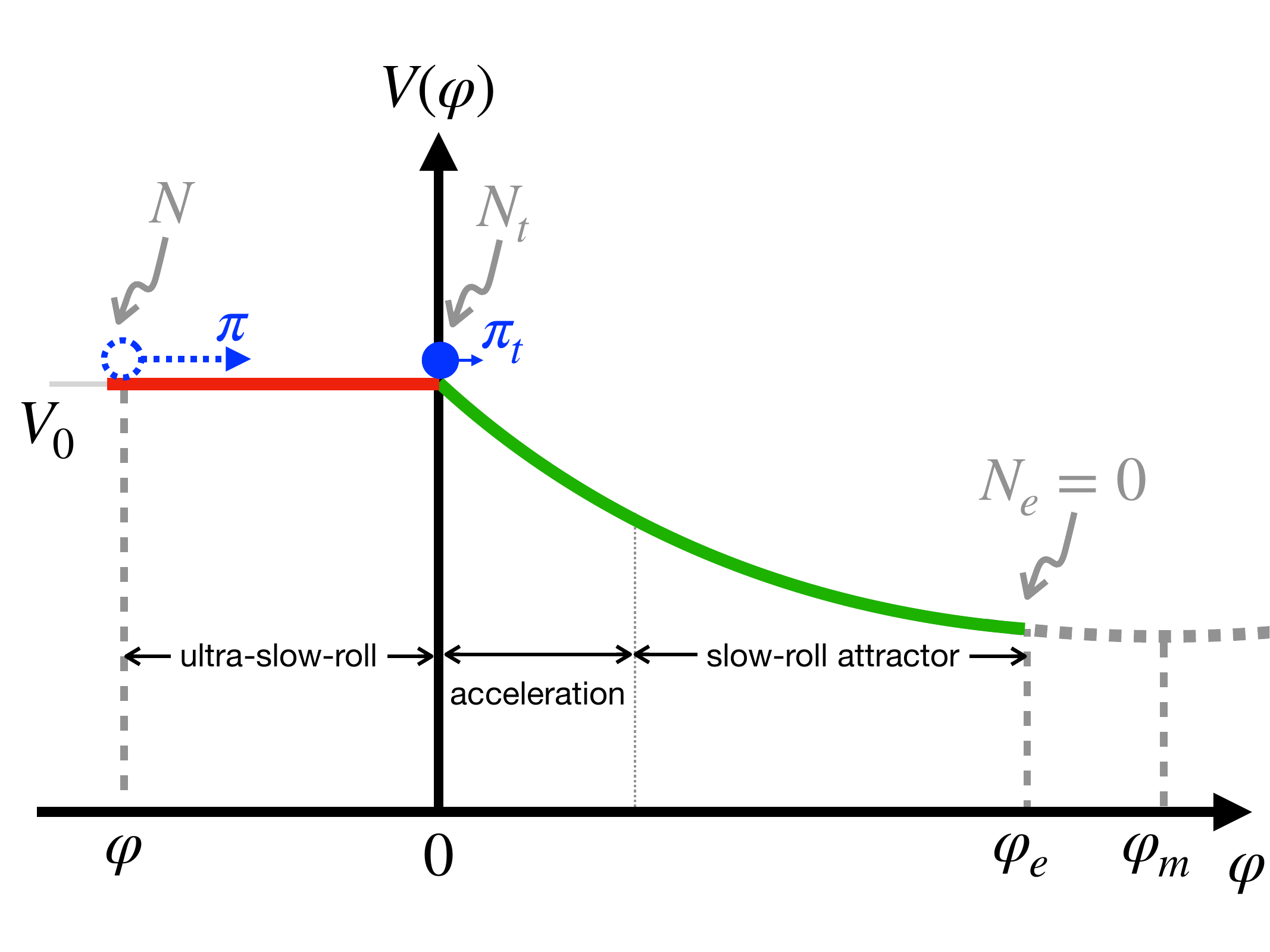}
\includegraphics[width=0.495\textwidth]{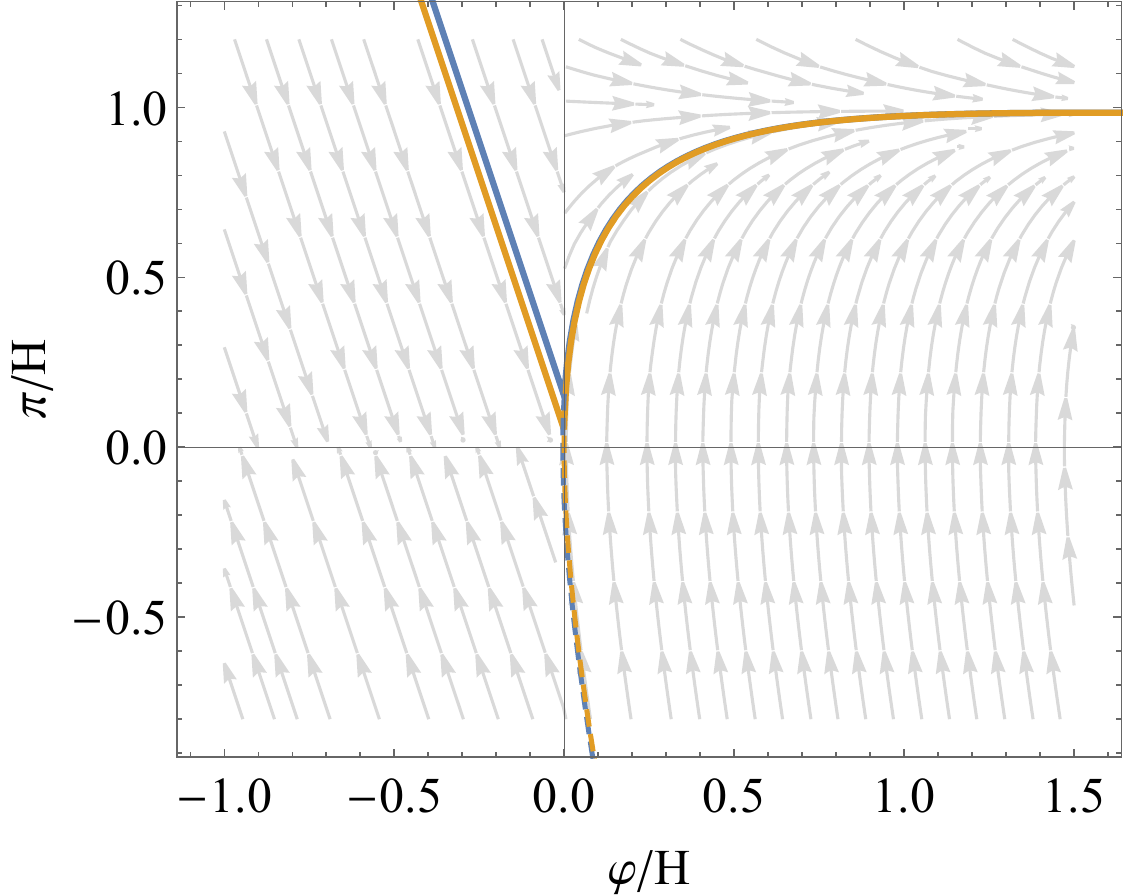}
\caption{The schematic potential (left panel) and the phase portrait (right panel) of the ``sharp transition'' from ultra-slow-roll to slow-roll, with $\eta'=0.01$, $\varphi_m=100$ (in the unit of $H$). For the fiducial (blue) and perturbed (orange) trajectories we set $\pi=9.15,~9.05$ at $\varphi=-3$, respectively. To show the difference of the trajectories clearly, $\delta\pi$ is exaggerated. Near the origin, the trajectories merge to a unique fast-roll trajectory with different $\pi$, which contribute negligibly to $\delta N$ in the subsequent evolution. The totel $\delta N$ comes mainly from the ultra-slow-roll stage, given by \eqref{eqn8:dN2}.}
\label{fig8:phase3}
\end{center}
\end{figure}

For the sharp transition case, as is shown in Fig. \ref{fig8:phase3}, the inflaton experiences an acceleration after the ultra-slow-roll stage, therefore the nearby trajectories, slightly shifted by the velocity difference $\delta\pi_{t}$ generated during the ultra-slow-roll stage, can also induce difference in $e$-folding number $\delta N$ in the later slow-roll stage. However, as $\pi_{t}\ll1$, trajectories with different $\pi_{t}$'s near the origin are almost on the same fast-roll trajectory in the slow-roll era for $\varphi>\varphi_{t}$, as the trajectories near the $\varphi$-axis are perpendicular to it. On this trajectory, the difference in $N$ caused by $\delta\pi_{t}$ is negligible, and $\varphi=\varphi_{t}$ is approximately an equal-$N$ surface. Following the discussion in Section \ref{sec8:usr}, we will arrive at the non-Gaussian relation \eqref{eqn8:dN2} straightforwardly.

\subsection{The curvaton scenario}\label{sec8:curvaton}
The curvaton scenario is a special two-field inflation model, of which the curvature perturbation is generated by the density perturbation of the curvaton field $\chi$ at its decay after inflation. This was first proposed in \cite{Moroi:2001ct,Enqvist:2001zp,Lyth:2002my} as an alternative method of generating curvature perturbation, which was usually referred to as the outcome of the quantum fluctuation of the inflaton field $\varphi$ during inflation. In the simplest curvaton models the power spectrum is nearly scale-invariant. To generate PBHs, the power spectrum should be enhanced on a specific scale, which can be realized in the axion-curvaton model \cite{Kasuya:2009up,Kawasaki:2012wr,Kawasaki:2013xsa,Ando:2017veq}, non-minimal curvaton scenario \cite{Pi:2021dft,Meng:2022low,Chen:2023lou}, or the multi-curvaton model \cite{Suyama:2011pu}. In the cuvaton scenario, it was shown that non-Gaussianity is inevitably enhanced compared to the slow-roll inflation, and the non-linear parameter is \cite{Bartolo:2003jx,Enqvist:2005pg,Sasaki:2006kq}\footnote{For simplicity, we only consider a quadratic potential for the curvaton $\chi$, and the decay is supposed to be instantaneous. For more details of these assumptions, see Ref. \cite{Sasaki:2006kq}.}  
\begin{equation}\label{eqn8:def:r}
f_\mathrm{NL}=\frac{5}{4r},\qquad
r\equiv\frac{3\Omega_{\chi,\mathrm{dec}}}{4-\Omega_{\chi,\mathrm{dec}}},
\end{equation}
where $r$ is roughly the cosmological density parameter of the curvaton $\chi$ at its decay. Apparently such a nominal $f_\mathrm{NL}$ is not enough to calculate the PBH abundance, as $f_\mathrm{NL}$ is always larger than 1. 


We will skip the model building and the dynamics of the curvaton field, but focus on its fully non-Gaussian curvature perturbation on the uniform-density slice, which is derived in Ref. \cite{Sasaki:2006kq} for a Gaussian curvaton field perturbation in the sudden decay approximation:
\begin{equation}\label{eqn8:zeta1}
\zeta=\ln\left(K^{1/2}\frac{1+\sqrt{ArK^{-3/2}-1}}{(3+r)^{1/3}}\right),
\end{equation}
where $r$ is defined in \eqref{eqn8:def:r}, and
\begin{align}
A&=\left(1+\frac{\delta\chi}{\chi}\right)^2,\quad\mbox{(on spatially-flat slicing)}\\
P&=(Ar)^2+\sqrt{(Ar)^4+(3+r)(1-r)^3},\\
K&=\frac{1}{2}\left[P^{1/3}+(r-1)(3+r)^{1/3}P^{-1/3}\right].
\end{align}
This fully non-linear curvature perturbation $\zeta$ can be used to calculate the PBH formation, following the procedure we show in Eq. \eqref{eqn8:routine2} \cite{Ferrante:2022mui,Ferrante:2023bgz,Gow:2023zzp}. 

The fully non-Gaussian curvature perturbation \eqref{eqn8:zeta1} seems complicated, but it has two interesting limiting cases for $r\ll1$ and $r\to1$, both of which have fruitful physical implications. In the former limit, the curvaton energy density is negligible at its decay, and 
\eqref{eqn8:zeta1} can be expanded for small $r$:
\begin{align}\label{eqn8:zeta2}
\zeta=\frac{r}{3} (A-1) -\frac{r^2}{9} (A-1)^2+\cdots=\frac{2}{3}r\frac{\delta\chi}{\chi}+\frac{r}{3}\left(1-\frac{4r}{3}\right)\left(\frac{\delta\chi}{\chi}\right)^2+\cdots,
\end{align}
where dots are higher order terms of $\mathcal{O}(r^3)$. 
We see that $\zeta$ has the familiar form of the quadratic local non-Gaussianity \eqref{eqn8:dN1}, with $\zeta_g=(2/3)r\delta\chi/\chi$ and $f_\text{NL}=5/(4r)$. Cubic and higher-order terms are suppressed when $r\ll1$, which makes it possible to truncate \eqref{eqn8:zeta2} at quadratic level even when $f_\mathrm{NL}\gg1$. Therefore, counterintuitively, such a quadratic expression \eqref{eqn8:zeta2} holds for large $f_\mathrm{NL}$, towards the limit of a $\chi^2$-distribution. 

The opposite limit $r\to1$ is more complicated and attracts much attention recently. In this limit, we simply have $P=2(Ar)^2$, $K=(Ar/2)^{2/3}$, and
\begin{align}\label{eqn8:zeta3}
\zeta=\frac{1}{3}\ln A=\frac{2}{3}\ln\left|1+\frac{\delta\chi}{\chi}\right|.
\end{align}
A simple derivation of \eqref{eqn8:zeta3} is as follows. As curvaton dominates the universe at its decay, the curvature perturbation on the uniform-curvaton slice $\zeta_\chi$ equals to the curvature perturbation on uniform density slice $\zeta$, which gives \cite{Pi:2021dft}
\begin{equation}
\zeta=\zeta_\chi(t,\mathbf{x})=\int^{\rho_\chi(t,\mathbf{x})}_{\rho_\chi(t)}\frac{\mathrm{d}\widetilde\rho_\chi}{3\widetilde\rho_\chi}
=\frac{1}{3}\ln\frac{\rho_\chi(t,\mathbf{x})}{\rho_\chi(t)}=\frac{2}{3}\ln\left|1+\frac{\delta\chi}{\chi}\right|,
\end{equation}
where in the last step we used $\rho_\chi=\frac{1}{2}m_\chi^2\chi^2$.
Therefore, in the $r\to1$ limit, the curvature perturbation $\zeta$ has a logarithmic dependence on the curvaton perturbation $\delta\chi$, which is similar to \eqref{eqn8:dN4}, but with a negative nonlinear parameter
\begin{equation}
f_\mathrm{NL}=-\frac{5}{4}.
\end{equation}
This means that the curvaton scenario can suppress the formation of PBHs in the $r\to1$ limit, which is a natural model to avoid the overproduction of PBHs when interpreting the nHz stochastic GWs as induced GWs \cite{Franciolini:2023pbf}.

\section{PBH abundance with non-Gaussian curvature perturbation}\label{sec8:abundance}

The non-Gaussian curvature perturbation discussed in the previous section may have a significantly different tail from the Gaussian tail in the PDF in the large $\mathcal{R}$ regime. 
As we commented, the quantum fluctuations of the inflaton field on subhorizon scales naturally provide a random variable $\delta \varphi$ which has a Gaussian distribution among all the Hubble patches, as the quantum fluctuation is Gaussian when the potential is approximately quadratic. 
In most of the cases, the Gaussian component of the curvature perturbation $\mathcal{R}_\mathrm{g}$ can be defined as proportional to the inflaton field perturbation $\delta\varphi$ which recovers the leading order Gaussian relation $\mathcal{R}_\mathrm{g}\equiv-(H/\dot\varphi)\delta\varphi$. As we have discussed in Section \ref{sec8:EPS}, the nonlinear relation $\mathcal{R}=\mathscr{F}(\mathcal{R}_\mathrm{g})$ can bring non-Gaussianities to the PDF of $\mathcal{R}$,  as is shown in \eqref{eqn8:R(Rg)}. Then the PBH abundance should be calculated following \eqref{eqn8:routine2}, by transferring the PDF $\mathbb{P}(\mathcal{R})$ to that of the linear compaction function $\mathbb{P}(\mathscr{C}_\ell)$. 

For instance, consider the following logarithmic relation of $\mathcal{R}$,
\begin{equation}\label{eqn8:dN2'}
\mathcal{R}=\mathscr{F}(\mathcal{R}_\mathrm{g})=\frac{1}{\lambda}\ln\left(1+\lambda\mathcal{R}_\mathrm{g}\right),\quad
\end{equation}
where $\mathcal{R}_\mathrm{g}$ is Gaussian. This formula represents ultra-slow-roll inflation ($\lambda=-3$), constant-roll inflation ($\lambda=\lambda_-$ defined in \eqref{eqn8:lambdapm}), and $r\to1$ limit of the curvaton scenario ($\lambda=3/2$). Knowing the PDF of $\mathcal{R}_\mathrm{g}$ and the Jacobian \eqref{eqn8:Jacobian}, the PDF of $\mathcal{R}$ is given by
\begin{align}\label{eqn8:exptail}
\mathbb{P}(\mathcal{R})
=\frac{e^{\lambda\mathcal{R}}}{\sqrt{2\pi}\sigma_{\mathcal{R}\mathrm{g}}}\exp\left[-\frac{\left(e^{\lambda\mathcal{R}}-1\right)^2}{2\lambda^2\sigma_{\mathcal{R}\mathrm{g}}^2}\right],
\end{align}
where $\sigma_{\mathcal{R}\mathrm{g}}$ is the root mean square of the Gaussian perturbation $\mathcal{R}_\mathrm{g}$, defined in \eqref{eqn8:sigmaR}.
For a negative $\lambda$ (including $\lambda=-3$), it goes as $\exp(\lambda\mathcal{R})$ when $\mathcal{R}\gg\mathcal{O}(0.1)$, so is called \textit{exponential tail} PDF. 
Such a PDF is less suppressed compared to the Gaussian tail, thus the PBH abundance will be greatly enhanced. Although $\mathbb{P}(\mathcal{R})$ is not directly used in the calculation of PBH mass function in the Press-Schechter-type formalism, it is straightforward to see the origin of such enhancement by the tail of $\mathbb{P}(\mathcal{R})$ when $\mathcal{R}\gg\sigma_{\mathcal{R}_\mathrm{g}}$. On the contrary, for a positive $\lambda$, the double exponential suppression is dominant, $\mathbb{P}\propto\exp(-c^2e^{\lambda\mathcal{R}})$, called \textit{Gumbel-distribution-like} PDF.

We will show how to calculate the PBH abundance for non-Gaussian curvature perturbation \eqref{eqn8:dN2'} in the Press-Schechter-type formalism, following \cite{Gow:2022jfb}. By using \eqref{eqn8:Cl-Gg}, the linear compaction function $\mathscr{C}_\ell$ can be written as 
\begin{align}
\mathscr{C}_\ell\equiv\frac{X}Y,\qquad\mathrm{with}\quad
X\equiv-\frac{4}{3}r\mathcal{R}'_\mathrm{g}(r),\quad
Y\equiv1+\lambda\mathcal{R}_\mathrm{g}.
\end{align}
The covariance matrix is $\boldsymbol\Sigma=\left(
\begin{matrix}
\sigma^2_{X} & \varrho\sigma_{X}\sigma_{Y}\\
\varrho\sigma_{X}\sigma_{Y} & \sigma^2_{Y}
\end{matrix}
\right)$, with
\begin{align}\label{eqn8:sx}
\sigma^2_{X}&=\langle XX\rangle=\left(-\frac{4}{3}\right)^2\int(kr)^2\mathcal{P}_{\mathcal{R}\mathrm{g}}(k)\left(\frac{\mathrm{d}j_0}{\mathrm{d}z}(kr)\right)^2\mathrm{d}\ln k,\\\label{eqn8:sxy}
\varrho\sigma_{X}\sigma_{Y}&=\langle XY\rangle=\left(-\frac{4}{3}\right)\lambda\int(kr)\mathcal{P}_{\mathcal{R}\mathrm{g}}(k)j_0(kr)\frac{\mathrm{d}j_0}{\mathrm{d}z}(kr)\mathrm{d}\ln k,\\\label{eqn8:sy}
\sigma^2_{Y}&=\langle YY\rangle=\lambda^2\int\mathcal{P}_{\mathcal{R}\mathrm{g}}(k)j_0^2(kr)\mathrm{d}\ln k,
\end{align}
where $\varrho$ is the correlation between $X$ and $Y$, $j_0(z)=\sin z/z$ is the zeroth order spherical Bessel function, and we assume a spherical $\zeta$ profile of
\begin{align}
\mathcal{R}_\mathrm{g}(r)=\frac{1}{\sqrt2\pi}\int\mathrm{d}k~kj_0(kr)\mathcal{R}_k,\qquad
\langle\mathcal{R}_k\mathcal{R}_p\rangle=\delta(k-p)\frac{2\pi^2}{k^3}\mathcal{P}_{\mathcal{R}\mathrm{g}}(k).
\end{align}
The joint PDF of $X$ and $Y$ becomes
\begin{align}\nonumber
\mathbb{P}(X,Y)&=\frac{1}{2\pi\sqrt{\mathrm{det}\boldsymbol\Sigma}}\exp\left[-\frac{1}{2}(X,\, Y-1)\boldsymbol{\Sigma}^{-1}\left(
\begin{matrix}
X \\ Y-1
\end{matrix}
\right)\right]\\\label{eqn8:P(X,Y)}
&=\frac{1}{2\pi\sigma_{X}\sigma_{Y}\sqrt{1-\varrho^2}}\exp\left(-\frac{X^2}{2\sigma^2_{X}}\right)
\exp\left[-\frac{1}{2\left(1-\varrho^2\right)}\left(\varrho\frac{X}{\sigma_{X}}-\frac{Y-1}{\sigma_Y}\right)^2\right],
\end{align}
which gives the PDF of $\mathscr{C}_\ell$,
\begin{align}\label{eqn8:P(Cl)}
\mathbb{P}(\mathscr{C}_\ell)=\int\mathrm{d}X\mathrm{d}Y~\mathbb{P}(X, Y)\delta\left(\mathscr{C}_\ell-\frac{X}Y\right)
=\int\mathrm{d}X~|Y|\cdot\mathbb{P}\left(X\to\mathscr{C}_\ell Y,Y\right).
\end{align}
The mass of the PBH obeys a power-law scaling from the critical collapse \cite{Choptuik:1992jv,Evans:1994pj,Koike:1995jm,Niemeyer:1997mt,Hawke:2002rf,Musco:2008hv}
\begin{align}\label{eqn8:M/MH}
\frac{M(\mathscr{C}_\ell)}{M_H}\sim K\left(\mathscr{C}-\mathscr{C}_\mathrm{th}\right)^\gamma
=K\left[\left(\mathscr{C}_\ell-\frac{3}{8}\mathscr{C}_\ell^2\right)-\mathscr{C}_\mathrm{th}\right]^\gamma,
\end{align}
where $\gamma\approx0.36$ and $K\sim1$. Solving this quadratic algebraic equation of $\mathscr{C}_\ell$, we have
\begin{align}\label{eqn8:Cl(M)}
\mathscr{C}_\ell=\frac{4}{3}\left(1-\sqrt{1-\frac{3}{2}\mathscr{C}_\mathrm{th}-\frac{3}{2}\left(\frac{M}{KM_H}\right)^{1/\gamma}}\right).
\end{align}
Another solution with $\mathscr{C}_\ell>4/3$ is discarded as it is for the Type II fluctuation. With \eqref{eqn8:M/MH}, to change the variable from $\mathscr{C}_\ell$ to $M$, we have
\begin{align}\label{eqn8:dCl/dlnM}
\frac{\mathrm{d}\mathscr{C}_\ell}{\mathrm{d}\ln M}=\frac{1}{\gamma}\frac{\mathscr{C}_\ell-\frac{3}8\mathscr{C}_\ell^2-\mathscr{C}_\mathrm{th}}{1-\frac{3}4\mathscr{C}_\ell}.
\end{align}
Therefore, substituting \eqref{eqn8:M/MH}, \eqref{eqn8:dCl/dlnM} together with \eqref{eqn8:P(Cl)} back into \eqref{eqn8:routine2}, we get the PBH mass function at the formation
\begin{align}
\beta(M)\mathrm{d}\ln M
&\approx K\frac{\left(\mathscr{C}_\ell-\frac{3}8\mathscr{C}_\ell^2-\mathscr{C}_\mathrm{th}\right)^{\gamma+1}}{\gamma\left(1-\frac{3}4\mathscr{C}_\ell\right)}\mathbb{P}(\mathscr{C}_\ell)\mathrm{d}\ln M,
\end{align}
where $\mathscr{C}_\ell$ should be expressed as a function of $M$, given by \eqref{eqn8:Cl(M)}.
The PBH mass function today is redshifted from the horizon reentry to equality, as the PBH energy density decays as $a^{-3}$ in the radiation dominated background $\sim a^{-4}$:
\begin{align}\label{eqn8:fPBH(M)}
f_\text{PBH}(M)&\approx3.81\times10^{8}\left(\frac{g_{*i}}{106.75}\right)^{-1/4}\left(\frac{h}{0.67}\right)^{-2}\left(\frac{M_\odot}{M_H}\right)^{1/2}
\left(\frac{M}{M_H}\right)^{-1/2}
\beta(M).
\end{align}
The PBH abundance today $f_\mathrm{PBH}=\int f_\mathrm{PBH}(M)\mathrm{d}\ln M$, is the integration of the mass function over all the mass range. 
For PBH as all the dark matter, $f_\mathrm{PBH}=1$.

\begin{figure}
\begin{center}
\includegraphics[width=0.48\textwidth]{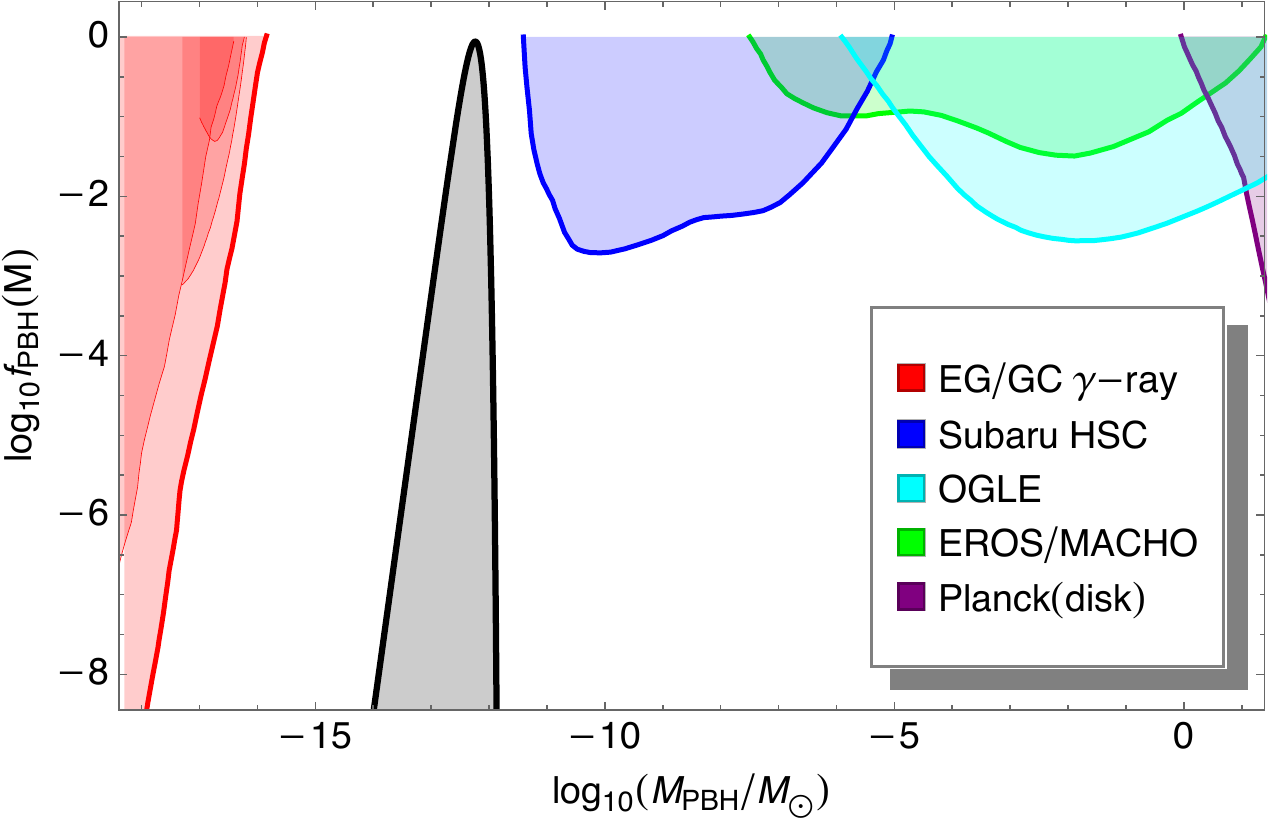}
\includegraphics[width=0.51\textwidth]{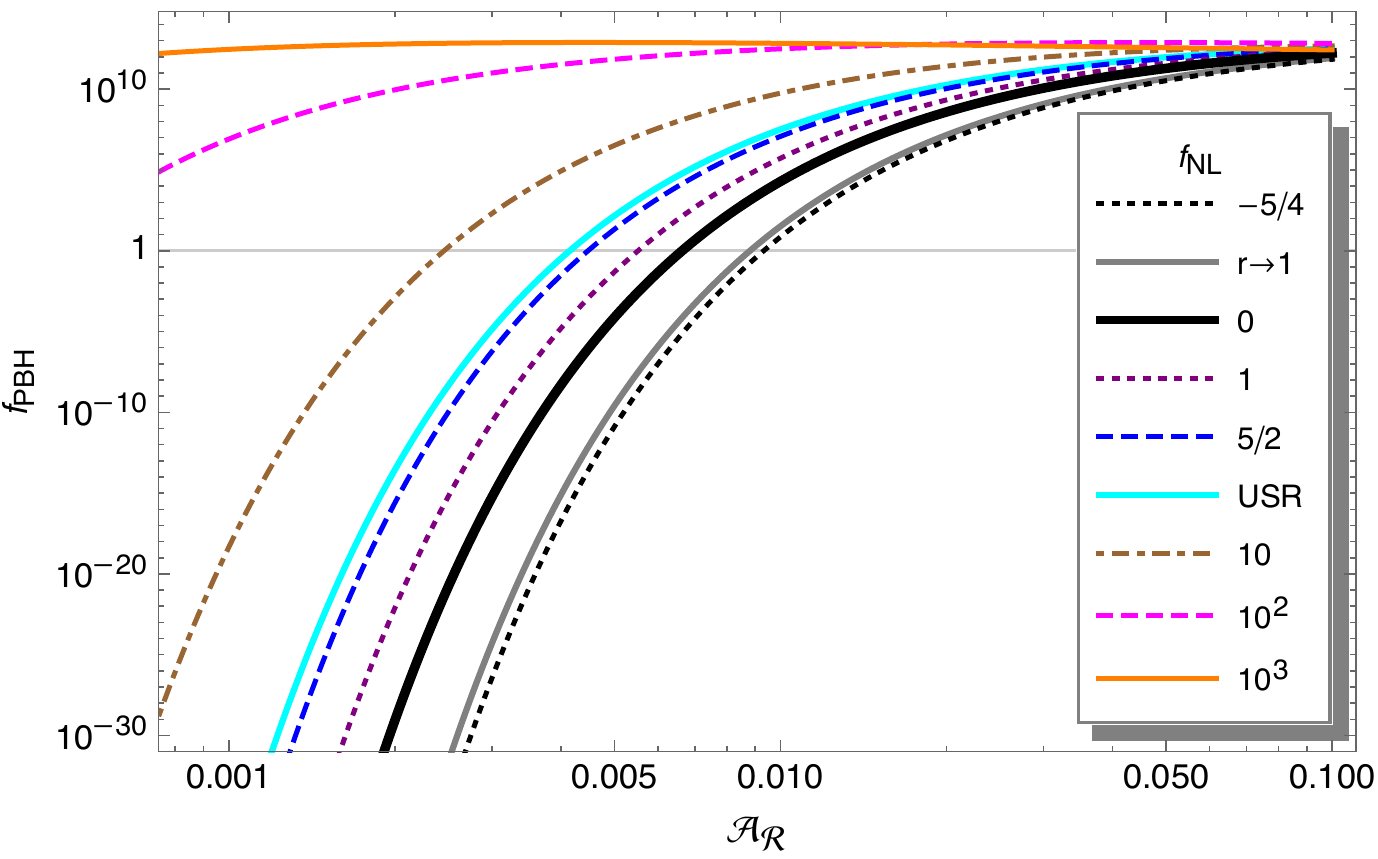}
\caption{
Left: The PBH mass function $f_\mathrm{PBH}(M)$ given by \eqref{eqn8:fPBH(M)} with a monochromatic power spectrum \eqref{eqn8:monoPR} in a ultra-slow-roll inflation with a sharp end, together with the some observational constraints. From left to right: extra-galactic $\gamma$-rays  \cite{Carr:2020gox}, the 511 keV line from galactic center \cite{DeRocco:2019fjq,Laha:2019ssq}, 21-cm signal from EDGES \cite{Mittal:2021egv}, Subaru HSC~\cite{Niikura:2017zjd}, EROS~\cite{Tisserand:2006zx}, OGLE \cite{Mroz:2024mse,Mroz:2024wag}, and CMB anisotropy with disk accretion \cite{Serpico:2020ehh}. 
The total power and central mass are $\mathcal{A_R}=4.151\times10^{-3}$ and $M_{k_*}=10^{-12}~M_\odot$, such that $\int f_\mathrm{PBH}(M)\mathrm{d}\ln M=1$.
Right: The total PBH abundance $f_\mathrm{PBH}$ as a function of the total power $\mathcal{A_R}$, for a monochromatic power spectrum \eqref{eqn8:monoPR}. From right to left, the curves are for $f_\mathrm{NL}=-5/4$ (black dotted), curvaton with $r\to1$ (gray), Gaussian case ($f_\mathrm{NL}=0$) (black thick), $f_\mathrm{NL}=1$ (purple dotted), $f_\mathrm{NL}=5/2$ (blue dashed), ultra-slow-roll inflation (cyan), $f_\mathrm{NL}=10$ (brown dot-dashed), $f_\mathrm{NL}=10^2$ (magenta dashed), and $f_\mathrm{NL}=10^3$ (orange), respectively.}
\label{fig8:fpbh}
\end{center}
\end{figure}

For simplicity, we only consider the monochromatic power spectrum 
\begin{align}\label{eqn8:monoPR}
\mathcal{P}_{\mathcal{R}\mathrm{g}}=\mathcal{A_R}\delta(\ln k-\ln k_*).
\end{align}
From \eqref{eqn8:sx} to \eqref{eqn8:sy},  we have 
\begin{align}\label{eqn8:sigma_X}
\sigma_X=\frac{4}{3}(k_*r)\mathcal{A}_\mathcal{R}^{1/2}\left|\frac{\mathrm{d}j_0}{\mathrm{d}z}(k_*r)\right|,\qquad
\sigma_Y=|\lambda|\mathcal{A}_\mathcal{R}^{1/2}\left|j_0(k_*r)\right|,
\end{align}
and $\varrho=\mathrm{sgn}(-\lambda j_0(k_*r)j_0'(k_*r))$. For simplicity, we choose $r_m$ to be the same as the Gaussian case, $k_*r_m\approx2.74$, which gives $\varrho=\mathrm{sgn}(\lambda)$. With $\varrho^2=1$, the PDF \eqref{eqn8:P(Cl)} reduces to one-dimensional, as $X$ and $Y$ are fully correlated. Together with $X=\mathscr{C}_\ell Y$, we have
\begin{align}
X=\frac{\sigma_X\mathscr{C}_\ell}{\sigma_X-\mathrm{sgn}(\lambda)\sigma_Y\mathscr{C}_\ell}, \qquad
Y=\frac{\sigma_X}{\sigma_X-\mathrm{sgn}(\lambda)\sigma_Y\mathscr{C}_\ell}. 
\end{align}
And the degenerated PDF of $\mathscr{C}_\ell$ becomes 
\begin{align}\label{eqn8:PDFexptail}
\mathbb{P}(\mathscr{C}_\ell)=\frac{1}{\sqrt{2\pi\sigma^2_X}}\left(\frac{\sigma_X}{\sigma_X-\mathrm{sgn}(\lambda)\sigma_Y\mathscr{C}_\ell}\right)^2\exp\left[-\frac{1}{2}\left(\frac{\mathscr{C}_\ell}{\sigma_X-\mathrm{sgn}(\lambda)\sigma_Y\mathscr{C}_\ell}\right)^2\right].
\end{align}
Apparently for $\lambda\to0$, it degenerates to the Gaussian PDF. Roughly speaking, the variance of PDF \eqref{eqn8:PDFexptail} increases for a negative $\lambda$ (\textit{i.e.} positive nominal $f_\mathrm{NL}$), which enhances PBH formation. As $\mathscr{C}_\ell$ increases, the PDF \eqref{eqn8:PDFexptail} finally approaches the Cauchy distribution $\mathbb{P}(\mathscr{C}_\ell)\propto\mathscr{C}_\ell^{-2}$, which, however, is far from the Type-I-PBH formation region of $0.872<\mathscr{C}_\ell<4/3$ \cite{Biagetti:2021eep}.
We calculated the PBH mass function $f_\mathrm{PBH}(M)$ for such a monochromatic power spectrum \eqref{eqn8:monoPR} in the ultra-slow-roll inflation ($\lambda=-3$), shown in the left panel of Fig. \ref{fig8:fpbh}. We choose a typical PBH peak mass of $M_{k^*}=10^{-12} M_\odot$, and normalize the amplitude of the power spectrum $\mathcal{A_R}$ to be $4.151077\times10^{-3}$, such that $f_\mathrm{PBH}=1$, \textit{i.e.} PBHs are all the dark matter. Interestingly, as is clearly seen in Fig. \ref{fig8:fpbh}, for the narrow peak case, the critical infrared scaling $f_\mathrm{PBH}(M)\propto M^{1+1/\gamma}\sim M^{3.78}$ which originates from the universal critical behavior is almost parallel to the constraints from the 21-cm signal which is proportional to $M^4$ \cite{Mittal:2021egv}. 
So the entire asteroid-mass window is open for narrow-peak power spectrum even with non-Gaussianity.

Another example is the perturbative series \eqref{eqn8:dN0}. Similarly we can define
\begin{align}
\mathscr{C}_\ell=XZ, \quad
X\equiv-\frac{4}{3}r\mathcal{R}'_\mathrm{g}(r),\quad
Z=1+\frac{6}{5}f_\mathrm{NL}\mathcal{R}_\mathrm{g}.
\end{align}
By the same method, for a narrow spectrum, we can get the PBH abundance $f_\mathrm{PBH}$ as a function of $\mathcal{A_R}$ for different $f_\mathrm{NL}$'s, given by \eqref{eqn8:routine2} with the PDF of $\mathscr{C}_\ell$ 
\begin{align}
\mathbb{P}\left(\mathscr{C}_\ell\right)=
\sum_{i=+,-}\frac{|Z_i|}{\sqrt{2\pi}|\sigma_XZ_i^2+\mathrm{sgn}(f_\mathrm{NL})\mathscr{C}_\ell\sigma_Z|}\exp\left(-\frac{\mathscr{C}_\ell^2}{2\sigma_X^2Z_i^2}\right),
\end{align}
where $\sigma_Z$ is given by \eqref{eqn8:sigma_X}, and 
\begin{align}
Z_\pm=\frac{1}{2}\pm\sqrt{\frac{1}{4}+\mathrm{sgn}(f_\mathrm{NL})\frac{\sigma_Z}{\sigma_X}\mathscr{C}_\ell},\qquad
\sigma_Z=\frac{6}{5}\mathcal{A}_\mathcal{R}^{1/2}\left|f_\mathrm{NL}j_0(k_*r)\right|.
\end{align}
When $f_\mathrm{NL}\to0$, the ``$+$'' branch goes back to Gaussian PDF, while the ``$-$'' branch disappears. Roughly speaking, for a positive $f_\mathrm{NL}$, the variance of $\mathbb{P}(\mathscr{C}_\ell)$ increases, which enhances PBH production. In the large-$f_\mathrm{NL}$ limit, the PDF reduces to $\chi^2$-distribution (a special form of gamma distribution):
\begin{align}
\mathbb{P}(\mathscr{C}_\ell)\approx\frac{1}{2\sqrt{2\pi\sigma_X\sigma_Z\mathscr{C}_\ell}}\exp\left(-\frac{\mathscr{C}_\ell}{2\sigma_X\sigma_Z}\right).
\end{align}
Besides its form, the key point is that in this limit only the product $\sigma_X\sigma_Z\propto\mathcal{A_R}f_\mathrm{NL}$ is relevant, and numerical integral shows that $f_\mathrm{PBH}=1$ requires $\mathcal{A_R}f_\mathrm{NL}\sim0.07$.

We can calculate the total abundance $f_\mathrm{PBH}$ as a function of $\mathcal{A_R}$ following \eqref{eqn8:routine2}, which is shown in the right panel of Fig. \ref{fig8:fpbh}, together with the result from ultra-slow-roll inflation and curvaton scenario with $r\to1$. As a positive $f_\mathrm{NL}$ enhances the PBH abundance, the $f_\mathrm{PBH}(\mathcal{A_R})$ curve moves upward for positive $f_\mathrm{NL}$'s, which makes the required $\mathcal{A_R}$ for $f_\mathrm{PBH}=1$ smaller, as we described in the introduction. Also, because of the higher order terms, the logarithmic relation \eqref{eqn8:dN2'} always generates more PBHs than the quadratic expansion with the same $f_\mathrm{NL}$, which requires a slightly smaller $\mathcal{A_R}$ for the same abundance of PBH. This is clearly shown for two cases: $f_\mathrm{NL}=5/2$ with ultra-slor-roll inflation, and $f_\mathrm{NL}=-5/4$ with curvaton scenario with $r\to1$.
The values of $\mathcal{A_R}$ required to realize PBH dark matter for $M_{k_*}=10^{-12} M_\odot$ are listed in Table \ref{tab8:AR}, which can be well fitted by
\begin{align}\label{eqn8:AR-fnl}
\mathcal{A_R}\approx\frac{6.6\times10^{-3}}{1+0.1f_\mathrm{NL}},
\end{align}
which recovers $\mathcal{A_R}f_\mathrm{NL}\to0.07$ when $f_\mathrm{NL}\gg1$.

\begin{table}
 \begin{tabular}{c c} 
   \hline
$f_\mathrm{NL}$ & $\mathcal{A_R}$ \\
  \hline
$-5/4$ & $9.328243\times10^{-3}$ \\
$r\to1$ & $8.808810\times10^{-3}$ \\
0 & $6.635506\times10^{-3}$ \\
0.1 & $6.498845\times10^{-3}$ \\
1 & $5.519888\times10^{-3}$ \\
  \hline
  \end{tabular}
  \hspace{2em}
   \begin{tabular}{c c} 
   \hline
$f_\mathrm{NL}$ & $\mathcal{A_R}$ \\
  \hline
5/2 & $4.479141\times10^{-3}$ \\
USR & $4.151077\times10^{-3}$ \\
10 & $2.456822\times10^{-3}$ \\
$10^2$ & $4.618021\times10^{-4}$ \\
$10^3$ & $5.690648\times10^{-5}$ \\
  \hline
  \end{tabular}
\caption{The total power $\mathcal{A_R}$ for a monochromatic power spectrum \eqref{eqn8:monoPR} needed to make PBH to be all the dark matter at $M_{k_*}=2\times10^{21}~\text{g}$, with different $f_\mathrm{NL}$'s. ``$r\to1$'' and ``USR'' stand for curvaton model with $r\to1$ and ultra-slow-roll inflation, described by a logarithmic $\mathcal{R}(\mathcal{R}_\mathrm{g})$ relation \eqref{eqn8:dN2'} with $\lambda=3/2$ ($f_\mathrm{NL}=-5/4$) and $\lambda=-3$ ($f_\mathrm{NL}=5/2$), respectively.}
\label{tab8:AR}
\end{table}


\section{Induced Gravitational Waves}\label{sec8:IGW}
The curvature perturbation can source gravitational waves via nonlinear coupling terms like scale-scalar-tensor. In this paper, we only briefly discuss the impact of the non-Gaussian curvature perturbation on the amplitude (via the normalization by the PBH abundance) as well as the shape of the induced GW spectrum. For a general discussion of scalar-induced GWs, see \cite{Witkowski:2021raz,Domenech:2021ztg} as well as Chapter 18 \cite{Domenech:2024kmh}.

For each polarization, the tensor perturbation obeys the following equation of motion \cite{Ananda:2006af,Osano:2006ew,Baumann:2007zm}
\begin{align}\label{eqn8:hk2}
h_\mathbf{k}''+2\mathcal{H}h_\mathbf{k}'+k^2h_\mathbf{k}=\int\frac{d^3p}{(2\pi)^{3/2}}e^{ij}_\mathbf{k}p_ip_j\left[2\Phi_\mathbf{p}\Phi_{\mathbf{k}-\mathbf{p}}+\frac{4\left(a\Phi_\mathbf{p}\right)'\left(a\Phi_{\mathbf{k}-\mathbf{p}}\right)'}{3(1+w)a^2\mathcal{H}^2}\right],
\end{align}
where $\Phi=\frac{2}{3}\mathcal{R}$ is the curvature perturbation in longitudinal gauge calculated in the radiation dominated era, and $e^{ij}_{\mathbf{k}}$ is one of the polarization tensor. 
After solving the equation of motion \eqref{eqn8:hk2}, we can calculate the energy spectrum of the GWs generated at the horizon reentry, and evaluate it until today. For simple spectral shape of $\mathcal{R}$ like $\delta$-function peak \cite{Kohri:2018awv}, lognormal peak \cite{Pi:2020otn}, flat plateau \cite{Saito:2009jt}, \textit{etc}., the GW spectrum can be calculated semi-analytically. Suppose the non-Gaussian curvature perturbation given by the local form \eqref{eqn8:R(Rg)}, $\mathcal{R}=\mathscr{F}(\mathcal{R}_\mathrm{g})$, can be expanded as
\begin{equation}\label{eqn8:dN0'}
\mathcal{R}=\mathcal{R}_g+F_\mathrm{NL}\mathcal{R}_g^2+G_\mathrm{NL}\mathcal{R}_g^3+H_\mathrm{NL}\mathcal{R}_g^4+I_\mathrm{NL}\mathcal{R}_g^5+\cdots
\end{equation}
where $F_\mathrm{NL}=(3/5)f_\mathrm{NL}$, $G_\mathrm{NL}=(9/25)g_\mathrm{NL}$, \textit{etc}. 
The GW spectrum today is given by a redshifted GW spectrum from the matter-radiation equality \cite{Cai:2018dig,Unal:2018yaa}
\begin{align}
\Omega_\text{GW}(f,\eta_0)h^2
=1.6\times10^{-5}\left(\frac{g_{*s}(\eta_k)}{106.75}\right)^{-1/3}\left(\frac{\Omega_{r,0}h^2}{4.1\times10^{-5}}\right)\Omega_\text{GW,eq}(f),
\end{align}
where
\begin{align}\nonumber
\Omega_\text{GW,eq}(k)
&=3\int^\infty_0\mathrm{d} v\int^{1+v}_{|1-v|}\mathrm{d} u\frac{1}{4u^2v^2}
\left[\frac{4v^2-(1+v^2-u^2)^2}{4uv}\right]^2\left(\frac{u^2+v^2-3}{2uv}\right)^4\\\nonumber
&\quad\cdot\left[\left(\ln\left|\frac{3-(u+v)^2}{3-(u-v)^2}\right|-\frac{4uv}{u^2+v^2-3}\right)^2+\pi^2\Theta\left(u+v-\sqrt3\right) \right]\\\nonumber
&\quad\cdot\left(\mathcal{P}_{\mathcal{R}\mathrm{g}}(uk)+F_\text{NL}^2\int^\infty_0 \mathrm{d}\nu\int^{1+\nu}_{|1-\nu|}\mathrm{d}\mu~\frac{\mathcal{P}_{\mathcal{R}\mathrm{g}}(\mu uk)\mathcal{P}_{\mathcal{R}\mathrm{g}}(\nu uk)}{\mu^2\nu^2}+\cdots\right)\\
\label{eqn8:OGW-NG}
&\quad\cdot
\left(\mathcal{P}_{\mathcal{R}\mathrm{g}}(vk)+F_\text{NL}^2\int^\infty_0 \mathrm{d}\lambda\int^{1+\lambda}_{|1-\lambda|}\mathrm{d}\rho~\frac{\mathcal{P}_{\mathcal{R}\mathrm{g}}(\lambda vk)\mathcal{P}_{\mathcal{R}\mathrm{g}}(\rho vk)}{\lambda^2\rho^2}+\cdots\right)+\cdots.
\end{align}
Here for simplicity we omit the higher order terms as well as the contribution from the connected part $\langle\mathcal{R}^4\rangle_c$, which are negligibly small when $F_\mathrm{NL}\lesssim10$. For the full expression, see Ref. \cite{Adshead:2021hnm} for quadratic expansion (up to $F_\mathrm{NL}$), Refs. \cite{Yuan:2023ofl,Li:2023xtl}  for cubic expansion (up to $G_\mathrm{NL}$), Ref. \cite{Perna:2024ehx} for quintic expansion \eqref{eqn8:dN0'} (up to $I_\mathrm{NL}$), and Ref. \cite{Abe:2022xur} for ultra-slow-roll inflation up to $\mathcal{A}_\mathcal{R}^4$.

The energy spectrum of the induced GW \eqref{eqn8:OGW-NG} can only be calculated for perturbative series like \eqref{eqn8:dN0'}. Fortunately, because the induced GWs mainly depends on the power spectrum of the curvature perturbation $\mathcal{P_R}$ which is at most $\sim10^{-2}$ for an enhanced peak, the higher order terms like $G_\mathrm{NL}\mathcal{R}^3$, $H_\mathrm{NL}\mathcal{R}^4$ contribute negligibly to the GW spectrum, as long as these coefficients $G_\mathrm{NL}$'s are not too large. Loose bounds for them are $G_\mathrm{NL}\ll\mathcal{O}(10^{2})F_\mathrm{NL}$, $H_\mathrm{NL}\ll\mathcal{O}(10^{4})F_\mathrm{NL}$, \textit{etc}., which are easy to be satisfied for most of the models.

There are two important examples. In the curvaton scenario with negligible curvaton at its decay, the perturbative series of \eqref{eqn8:dN0} is valid even if $f_\mathrm{NL}\gg1$, as the higher order terms ($G_\mathrm{NL}\mathcal{R}^2$ \textit{etc}.) are always strongly suppressed. In the ultra-slow-roll inflation with a sharp end, $\mathcal{R}(\mathcal{R}_\mathrm{g})$ is given by \eqref{eqn8:dN2'}, of which the nonlinear parameters are
\begin{equation}
F_\mathrm{NL}=3/2,\quad G_\mathrm{NL}=3, \quad H_\mathrm{NL}=27/4, \quad\cdots.
\end{equation}
This surely guarantees that higher-order contributions are negligible, which was shown explicitly up to $\mathcal{A}_\mathcal{R}^4$ in Ref. \cite{Abe:2022xur}.

\begin{figure}[htbp]
\includegraphics[width=0.51\textwidth]{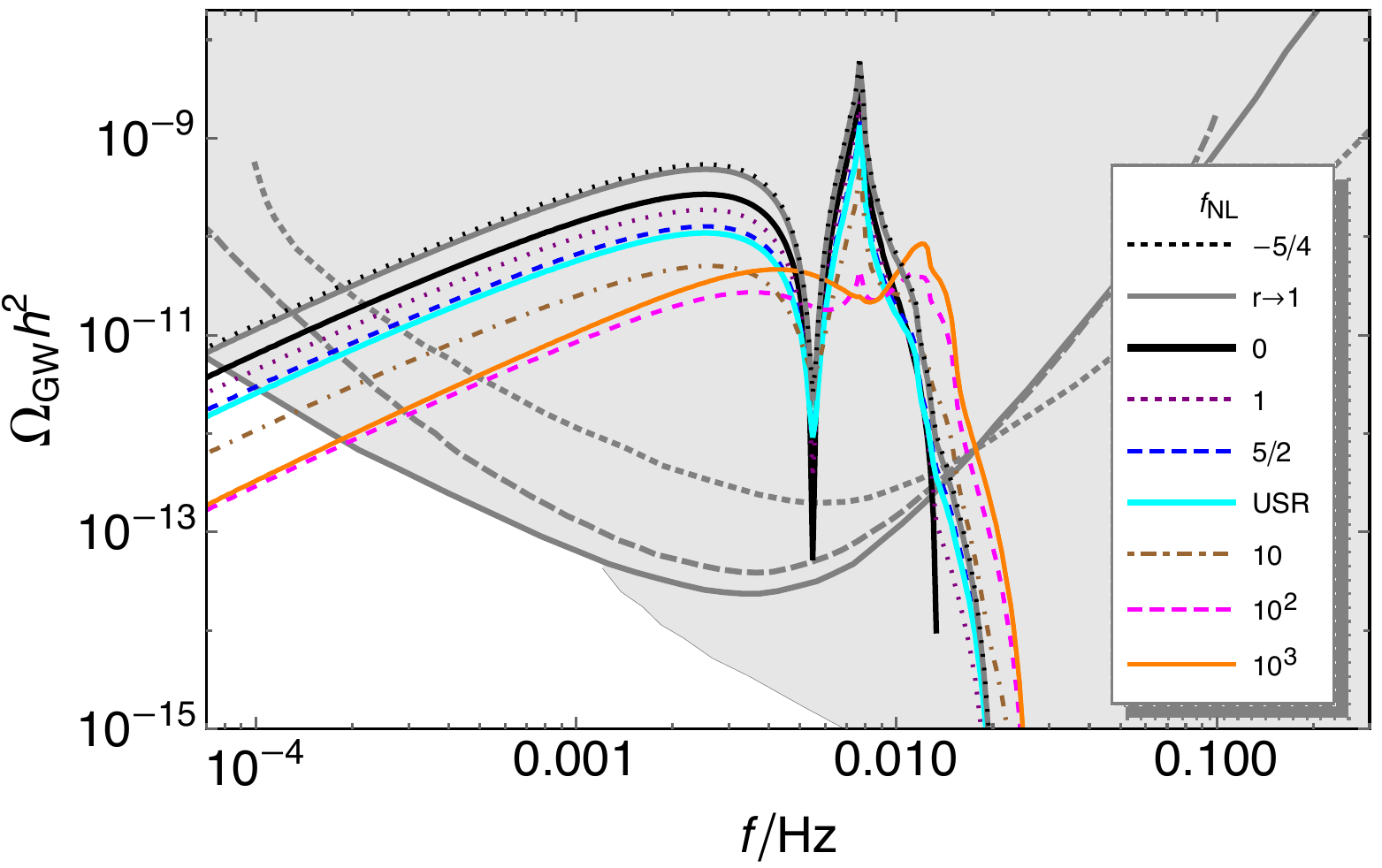}\!\!
\includegraphics[width=0.48\textwidth]{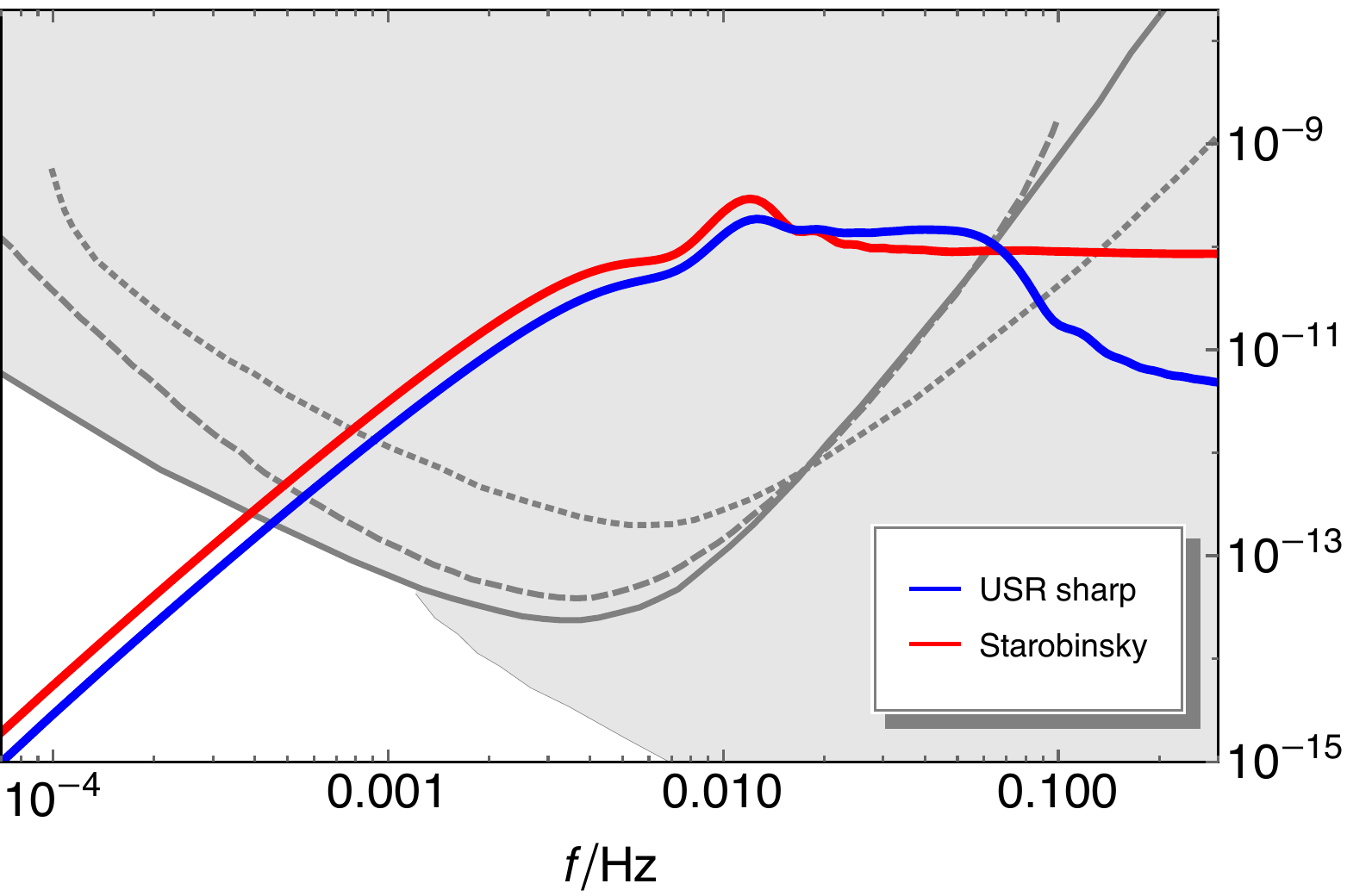}
\caption{
Left: The spectrum of GWs induced by the enhanced narrow-peak power spectrum \eqref{eqn8:monoPR}. The amplitude $\mathcal{A_R}$ for different parameters follows Table \ref{tab8:AR}, such that PBHs $\sim 10^{-12} M_\odot$ are all the dark matter, shown in Fig.\ref{fig8:fpbh}. From right to left, the orange, magenta dashed, brown dot-dashed, cyan thick, blue dashed, purple dotted, black thick, gray, and black dotted curves are for $f_\mathrm{NL}=10^3$, $f_\mathrm{NL}=10^2$, $f_\mathrm{NL}=10$, ultra-slow-roll, $f_\mathrm{NL}=5/2$, $f_\mathrm{NL}=1$, Gaussian case ($f_\mathrm{NL}=0$), curvaton scenario with $r\to1$, and $f_\mathrm{NL}=-5/4$, respectively. The curves are drown by \cite{adshead_2023_8176108}, following \cite{Adshead:2021hnm}, together with the designed power-law integrated sensitivity curve of space-borne interferometers LISA (gray thick) \cite{Bartolo:2016ami}, Taiji (gray dashed) \cite{Wang:2021njt}, TianQin (gray dotted)~\cite{Liang:2021bde}, and DECIGO (gray thin) \cite{Kawamura:2006up,Kawamura:2011zz}. Right: Two more realistic signals, the spectrum of GWs induced by ultra-slow-roll inflation with a sharp transition to slow-roll (blue), and a smooth transition to slow-roll, \textit{i.e.} Starobinsky's linear potential model (red). The right panel was drawn by \eqref{eqn8:OGW-NG} with the help of the SIGWfast \cite{Witkowski:2022mtg} package.}
\label{fig8:IGW}
\end{figure}

According to the current observational constraints, asteroid mass PBHs can serve as all the dark matter, of which the scalar induced GWs are in the milliHz to deciHz band, to be explored by the space-borne interferometers. As an concrete example, we calculate the induced GW spectra for monochromatic power spectra \eqref{eqn8:monoPR} with different $f_\mathrm{NL}$'s, normalized such that PBHs of $M_{k_*}\approx 10^{-12} M_\odot$ are all the dark matter. The resultant induced GWs peak at $k_*=6.73\times10^{-3}$ Hz, which are shown in Fig. \ref{fig8:IGW} together with the power-law integrated sensitivity curves \cite{Thrane:2013oya} of the space-borne interferometers LISA~\cite{AmaroSeoane:2012km,AmaroSeoane:2012je,Audley:2017drz,Barausse:2020rsu}, Taiji~\cite{Guo:2018npi}, TianQin~\cite{Luo:2015ght}, and DECIGO \cite{Schmitz:2020syl}. The induced GW from Gaussian curvature perturbation with a monochromatic power spectrum displays an infrared scaling of $\propto k^2$, a dip of zero at $\sqrt{2/3}k_*$, a resonant peak of logarithmic divergence at $(2/\sqrt3)k_*$, and a sharp cutoff beyond $2k_*$ \cite{Kohri:2018awv}. This infrared scaling is a consequence of the infinitesimal scale of the source \cite{Cai:2019cdl}, which is preserved even in the non-Gaussian case \cite{Adshead:2021hnm}. As we commented before, positive $f_\mathrm{NL}$ can enhance the PBH formation, which requires less $\mathcal{A_R}$ for a fixed PBH abundance (\textit{e.g.} $f_\mathrm{PB H}=1$), thus smaller GW spectrum is displayed. However, when $f_\mathrm{NL}\lesssim10$, the suppression is within one order of magnitude, which makes the amplitude and shape of the GW spectrum similar to the Gaussian one and difficult to be distinguished. In this case, the GWs beyond the cutoff frequency $2k_*$ is purely from the non-Gaussian part, which is a smoking gun of primordial non-Gaussianity \cite{Cai:2018dig,Abe:2022xur}. Although these contributions are tiny compared with the main peak, they could be probed by the deci-hertz GW detector like BBO ($0.1$ to 1~Hz)~\cite{Crowder:2005nr,Corbin:2005ny} 
or DECIGO ($10^{-2}$ to 1~Hz)~\cite{Kawamura:2006up,Kawamura:2011zz}.

As $f_\mathrm{NL}$ increases, the main peak contributed by the Gaussian part decreases due to the less requirement of $\mathcal{A_R}$ because of the enhanced productivity of PBHs. However, the non-Gaussian contribution increases slighly, and finally dominates the GW spectrum and changes its shape significantly. The non-Gaussian source is roughly a convolution of two spectra with different wavenumbers, which fills the dip at $\sqrt{2/3}k_*$. 
Also, $\Omega_\mathrm{GW}^\mathrm{(NG)}$ has a mild bump instead of a sharp peak at $\sim\sqrt3 k_*$, of which the amplitude is $\sim F_\mathrm{NL}^4\mathcal{A}_\mathcal{R}^4$. It becomes visible by LISA for $f_\mathrm{NL}\gtrsim20$. 

A very important issue is that when PBHs constitute all the dark matter, $\Omega_\mathrm{GW}$ has a lower bound which is detectable by the space-borne interferometers \cite{Cai:2018dig}. 
To see it clearly, note that the peak value of $\Omega_\mathrm{GW}$ is given by the maximum of the three peaks contributed by $\mathcal{O}(\mathcal{A}_\mathcal{R}^2)$-, $\mathcal{O}(\mathcal{A}_\mathcal{R}^3f_\mathrm{NL}^2)$-, and $\mathcal{O}(\mathcal{A}_\mathcal{R}^4f_\mathrm{NL}^4)$-terms:
\begin{align}\nonumber
\Omega_\mathrm{GW,peak}h^2&\approx1.6\times10^{-5}\mathrm{max}\Big[6.4\mathcal{A}_\mathcal{R}^2,~3.7\mathcal{A}_\mathcal{R}^3F_\mathrm{NL}^2,~3.9\mathcal{A}_\mathcal{R}^4F_\mathrm{NL}^4\Big],\\\nonumber
&\approx\mathrm{max}\left[\frac{4.5\times10^{-9}}{(1+0.1f_\mathrm{NL})^2},~\frac{6.1\times10^{-12}}{(1+0.1f_\mathrm{NL})^3}f_\mathrm{NL}^2,~\frac{1.5\times10^{-14}}{(1+0.1f_\mathrm{NL})^4}f_\mathrm{NL}^4\right].
\end{align}
In the first expression, the numerical coefficient of the Gaussian contribution (``6.4'') comes from Ref. \cite{Pi:2020otn} with smoothing scale $\Delta=0.01$, while the coefficients of the $F_\mathrm{NL}^2$-term (``3.7") and $F_\mathrm{NL}^4$-term (``3.9") come from Ref. \cite{Adshead:2021hnm} (also with a width of $\Delta=0.01$). In the second equality, we use \eqref{eqn8:AR-fnl} for PBH as all the dark matter. From the above fitting formula, it is easy to see that $\Omega_\mathrm{GW,peak}h^2$ has a minimum of $\Omega_\mathrm{GW,peak}^\text{(min)}h^2\sim8.4\times10^{-11}$ when $f_\mathrm{NL}\sim63$. After that, $\Omega_\mathrm{GW,peak}h^2$ only increase slightly, and finally approaches a constant of $1.1\times10^{-10}$.
This minimum is still well above the sensitivity curves of space-borne interferometers. Therefore, the detectability of scalar-induced GWs is robust against the nonlinear parameter $f_\mathrm{NL}$, $\lambda_-$ (see \eqref{eqn8:lambdapm} and \eqref{eqn8:dN4}), or $r$ (see \eqref{eqn8:def:r} and \eqref{eqn8:zeta1}),  which made it an important scientific goal of LISA \cite{Barausse:2020rsu,LISA:2022kgy,LISACosmologyWorkingGroup:2022jok,LISACosmologyWorkingGroup:2023njw}, Taiji \cite{Ren:2023yec}, and TianQin \cite{Liang:2021bde} to probe the PBH dark matter. 

We should warn that the $\delta$-function peak we considered above is not physical. For more realistic models, the enhancement and decay of the power spectrum around the peak are milder, usually obeying power-law. Especially, in the typical model of ultra-slow-roll inflation, $\mathcal{P}_\mathcal{R}(k)$ increases as $\propto k^4$  before it reaches the maximum \cite{Byrnes:2018txb}. There might be modulated oscillations of period $\sim\pi$ if the transition from slow-roll to ultra-slow-roll is abrupt \cite{Starobinsky:1992ts,Biagetti:2018pjj,Cole:2022xqc,Pi:2022zxs}. The UV behavior of the power spectrum depends on the shape of the second slow-roll potential, usually displaying another power-law $\mathcal{P_R}\propto k^{n_{s,2}-1}$. If the ultra-slow-roll stage ends sharply, the power spectrum drops off a little bit with new modulated oscillations with a larger period of $\sim\pi e^{N_\mathrm{USR}}$ \cite{Pi:2022zxs}. According to our discussion in Section \ref{sec8:logdual}, these oscillations mark the sharp end of ultra-slow-roll stage, which are accompanied with the logarithmic relation \eqref{eqn8:dN2}. Based on the analytical expression of $\mathcal{P}_\mathcal{R}(k)$ as well as the statistical properties, we can draw the GW spectrum associated with a fixed PBH abundance. Currently, there is no satisfactory method of calculating PBH abundance for such a broad spectrum with non-Gaussianity, so we choose the same parameters as in the narrow peak case (marked ``USR'' in Table \ref{tab8:AR}), aiming to demonstrate the feature but not the exact amplitude of the induced GW spectrum in the ultra-slow-roll inflation. This is shown in the right panel of Fig. \ref{fig8:IGW}. For comparison, we also draw the induced GW from Starobinsky's linear potential model (with the narrow-peak amplitude), which represents the smooth end of ultra-slow-roll stage with negligible non-Gaussianity. We can see that similar to the $\delta$-function peak case, non-Gaussianity enhances the UV part of the induced GW spectrum, which makes the primary peak of the GW spectrum obscure. A drop off with modulated oscillations can be recognized as a smoking gun for the non-Gaussianity in the ultra-slow-roll inflation shown in \eqref{eqn8:dN2}, which are detectable by DECIGO and BBO when PBHs constitute all the dark matter. 

When the transition from ultra-slow-roll to the second slow-roll stage is smooth, the curvature perturbation is dominated by the $\delta N$ contributed by the second slow-roll stage, which has the normal consistency relation $f_\mathrm{NL}=(5/12)(1-n_{s,2})$ with $n_{s,2}$ the UV tilt of $\mathcal{P_R}(k)$. In such induced GW spectra, there will be no modulated oscillations with period of $\sim\pi e^{N_\mathrm{USR}}$, but $f_\mathrm{NL}$ is determined by the UV slope of the GW spectrum, as $\Omega_\mathrm{GW}(k)\sim k^{2(n_{s,2}-1)}$. This can also be used to probe non-Gaussianity \cite{Atal:2021jyo}. 


In summary, it is clearly shown in Fig. \ref{fig8:fpbh} and Fig. \ref{fig8:IGW} that the dependence of the induced GW amplitude on the non-Gaussianity is mild once the PBH abundance is fixed. However, the non-Gaussian impact on the spectral shape could be detected in the future. 


\section{Conclusion and Discussion}\label{sec8:conclusion}

In this paper we review the impact of the primordial non-Gaussinity in the curvature perturbation on the PBH formation as well as on the generation of scalar-induced GWs. We discussed some concrete inflation models to realize these interesting non-Gaussian curvature perturbations, and demonstrate how to calculate the PBH abundance as well as the induced GWs when non-Gaussianity is taken into account. Positive/negative non-Gaussianity can greatly enhance/suppress the PBH abundance, as they change the tail of the PDF significantly. However, the induced GW is mainly determined by the power spectrum, thus non-Gaussianity can not significantly change the induced GW spectrum. When fixing the PBH abundance, positive/negative non-Gaussianity only mildly suppresses/enhances the induced GWs. Therefore, the amplitude of the induded GW spectrum in the presence of abundant PBHs is a relatively robust prediction, which is an important scientific goal for many current and future experiments. 



We only focus on the simplest local-type non-Gaussianity, which gives $\mathcal{R}(\mathcal{R}_\mathrm{g})$ as a function of Gaussian variable $\mathcal{R}_\mathrm{g}$. Scale dependence of such a relation is not considered here \cite{Passaglia:2018ixg,Ragavendra:2021qdu,Ozsoy:2021pws,Ragavendra:2023ret,Tasinato:2023ioq}. Especially, various types of non-Gaussianities like equilateral, orthogonal, or folded shapes are also important, not only on the abundance but also on the clustering \cite{Shandera:2012ke,Young:2015cyn,Tada:2015noa,Young:2019gfc,Suyama:2019cst,Matsubara:2019qzv,Ragavendra:2020sop,DeLuca:2021hcf}. Recently, Ref. \cite{Matsubara:2022nbr} studied the bispectrum and trispectrum in the PBH formation systematically, and shows that in the narrow power spectrum case, the nonlinear parameter $f_\mathrm{NL}$ entering the calculation of PBH formation should be a linear combination of $f_\mathrm{NL}$ of local, equilateral, folded, and orthogonal shapes:
\begin{align}
f_\mathrm{NL}^\mathrm{eff}\equiv f_\mathrm{NL}^\mathrm{loc}-3f_\mathrm{NL}^\mathrm{eq}+3f_\mathrm{NL}^\mathrm{fo}-9f_\mathrm{NL}^\mathrm{ort},
\end{align}
which implies that other shapes of non-Gaussianity should also be taken into account when calculating the PBH abundance. Such non-Gaussianities are important for inflation model with non-canonical kinetic terms, like for instance 
k-essence 
or G-inflation \cite{Lin:2020goi,Yi:2020cut,Zhang:2020uek,Gao:2021vxb,Choudhury:2023kdb}. 

Our starting point is the $\delta N$ formalism, which assumes that the Hubble patches we study are far enough from each other, such that each patch can be seen as a local Friedmann universe where the field value (and other perturbed quantities) takes a random initial value and evolves independently in the later stage. This is called \textit{seperate universe approach}, which works quite well in the slow-roll regime. However, in the beginning of ultra-slow-roll stage (marked by $\varphi_\mathrm{i}$), the gradient terms of both growing mode and decaying mode are rather important, which give rise to the well-known $k^4$ growth \cite{Leach:2001zf,Byrnes:2018txb,Cole:2022xqc}. Such a $k$-dependence cannot be recovered by seperate universe approach and the classical $\delta N$ formalism based on it \cite{Jackson:2023obv}. Fortunately, such a failure is only serious for $\varphi\lesssim\sqrt\pi\varphi_\mathrm{i}$ \cite{Domenech:2023dxx}, which makes the seperate universe approach and the $\delta N$ formalsm still valid at the peak of the enhanced power spectrum, which is located at $\varphi_\mathrm{p}\approx\pi\varphi_\mathrm{i}$ \cite{Pi:2022zxs}. 
Apparently, if we extend the $\delta N$ formalism to include the spatial curvature on the initial slice taken slightly after horizon exit, the leading order correction of the gradient expansion can be taken into account \cite{Artigas:2024ajh}.

We did not discuss the interesting case of negative non-Gaussianity, which is connected to Type II PBHs. Simple estimate shows that negative non-Gaussianity will significantly suppress the PBH abundance, which was calculated in a perturbative way \cite{Young:2013oia}. For fully nonlinear curvature perturbation with a logarithmic dependence on the Gaussian variable, the PDF of $\mathcal{R}$ displays a Gumbel-like distribution \cite{Pi:2022ysn} (also called double exponential suppression \cite{Hooshangi:2023kss}):
\begin{align}\label{eqn8:gumbel}
\mathbb{P}(\mathcal{R})
\sim \exp\left(-c^2e^{2\lambda_-\mathcal{R}}\right).
\end{align}
This can be realized, for instance, in the $r\to1$ limit of the curvaton scenario which corresponds to $\lambda_-=3/2$. Direct calculation shows that it can significantly suppress the PBH formation. This suppression was utilized to avoid the overproduction of sub-solar-mass PBHs when interpreting the nHz stochastic GW background as the scalar-induced GWs \cite{Franciolini:2023pbf,Liu:2023ymk,Choudhury:2023fwk}. Recent study by the theory of peaks indicates that the PBH formation for negative $f_\mathrm{NL}$ is quite difficult \cite{Kitajima:2021fpq}, and numerical simulations show that there is a lower bound $f_\mathrm{NL}\gtrsim-1.2$ below which the Type I PBHs can not form, based on a monochromatic power spectrum \cite{Escriva:2022pnz}. Around this regime the profiles of the density peaks become complicated, and the Type II PBH are crucial \cite{Uehara:2024yyp}, which we did not touch here.


In many models, the non-attractor behavior is excited by the sudden change of the slop of the potential, which itself is continuous. This actually means that the change of slope is realized in an energy scale much larger than $H$, \textit{i.e.} $V''\gg H^2$ at $\varphi_{t}$. Another class of models display a discontinuity or a bump in the potential, which causes a loss of kinetic energy. Large non-Gaussianity can also be generated in such a case, which gives a highly asymmetric PDF of $\mathcal{R}$ \cite{Cai:2021zsp,Cai:2022erk,Briaud:2023eae,Kawaguchi:2023mgk}. If the initial velocity is not enough, the inflaton may not overshoot the step/bump in some Hubble patches. Recent study shows that such a Hubble patch finally collapse to a PBH, as the energy density in the surrounding area become smaller than this patch and starts to compress it. When $f_\mathrm{NL}\gtrsim2.6$, the PBHs from such bubbles are dominant \cite{Escriva:2023uko,Caravano:2024tlp}. 

In principle, for non-quadratic potential, the equation of motion for the inflaton cannot be solved analytically, and the analysis based on quadratic potentials in Section \ref{sec8:logdual} cannot apply. However, inspired by the logarithmic relation of $\mathcal{R}(\mathcal{R}_\mathrm{g})$ and its exponential-tail PDF \eqref{eqn8:exptail}, it is reasonable to conjecture a more general PDF of $\mathcal{R}$ directly \cite{Nakama:2016kfq,Hooshangi:2021ubn,Hooshangi:2023kss}. For instance, a simple guess is
\begin{equation}\label{eqn8:heavytail}
\mathbb{P}(\mathcal{R})\propto\exp\left(-c\left|\mathcal{R}\right|^p\right),
\end{equation}
where $c$ is a positive constant. The PDF with $0<p<1$ is called heavy tailed, which is more effective in generating PBHs than the exponential-tail PDF. This extra efficiency is needed to generate enough supermassive PBHs to seed early galaxies, as the power spectrum of $\mathcal{R}$ on that scale is strongly constrained by $\mu$- and $y$-distortions, $\mathcal{P_R}\lesssim10^{-5}$ \cite{Kohri:2014lza,Nakama:2017xvq,Carr:2018rid,Atal:2020yic,Liu:2022bvr,Gouttenoire:2023nzr}. Such a heavy-tail PDF can arise from self-interacting curvaton scenario \cite{Hooper:2023nnl}, while a lighter tail of $p=3/2$ can be generated in the non-perturbative regime of inflation model with a $\dot{\mathcal{R}}^4$ interaction \cite{Celoria:2021vjw}. 

Up to now, all the discussions above are based on the classical $\delta N$ formalism, which allows us to divide the field value $\varphi$ in a Hubble patch into ``background'' $\varphi_0$ and ``perturbation'' $\delta\varphi$. Such a split becomes difficult (if not completely impossible) when the perturbations are large and overwhelm the background value. As we commented, this happens when the inflaton is nearly static on a flat plateau due to the Hubble friction, and quantum diffusion must be taken into account \cite{Starobinsky:1986fx,Starobinsky:1982ee,Starobinsky:1994bd}. Although the exponential-tail PDF like \eqref{eqn8:exptail} appears as the leading order of the PDF in the semi-classical limit, more complicated and bizarre PDF arises in the quantum regime \cite{Bullock:1996at,Vennin:2015hra,Pattison:2017mbe,Ezquiaga:2019ftu,Vennin:2020kng,Figueroa:2020jkf,Pattison:2021oen,Figueroa:2021zah,Animali:2022otk,Ezquiaga:2022qpw}. This is the topic of the next chapter \cite{Vennin:2024yzl}.

\section*{Acknowledgement}
I would like to thank Diego Cruces, Guillem Dom\`{e}nech, Jaume Garriga, Cristiano Germani, Cristian Joana, Misao Sasaki, Teruaki Suyama, Takahiro Tanaka, Dong-Gang Wang, and Jianing Wang for their comments and discussions, and to thank Ryoto Inui and Ao Wang for helping me draw Fig \ref{fig8:IGW}. I also thank the hospitality of YITP, Kavli IPMU, IBS-CTPU, and KIAS during my visit when this paper was finalized, and thank all the participants of the 2nd International Workshop on Gravitational Waves and the Early Universe for the intriguing discussions. This work is supported in part by the National Key Research and Development Program of China Grant No. 2021YFC2203004, by Project No. 12475066 and No. 12447101 of the National Natural Science Foundation of China, by the JSPS Grant-in-Aid for Early-Career Scientists No. JP20K14461, by JSPS KAKENHI No. JP24K00624, and by the World Premier International Research Center Initiative (WPI Initiative), MEXT, Japan.

\bibliography{NG-PBH.bib}

\end{document}